\documentclass[%
 reprint,
 amsmath,amssymb,
 aps,
]{revtex4-2}

\usepackage{graphicx}
\usepackage{dcolumn}
\usepackage{bm}
\usepackage{caption}
\usepackage{subcaption}
\usepackage{algorithmic}
\usepackage[ruled,vlined]{algorithm2e}

\begin{document}

\preprint{APS/123-QED}

\title{Adaptive phase-retrieval stochastic reconstruction with correlation functions: 3D images from 2D cuts}

\author{Aleksei Cherkasov}
 \email{cherkasov.am@phystech.edu}

\affiliation{%
 Moscow Institute of Physics and Technology, 9 Institutskiy per., Dolgoprudny, Moscow Region, 141701, Russian Federation }%
 
\author{Andrey Ananev}%

\affiliation{%
 Moscow Institute of Physics and Technology, 9 Institutskiy per., Dolgoprudny, Moscow Region, 141701, Russian Federation }%

\author{Marina Karsanina}
\affiliation{Schmidt Institute of Physics of the Earth of Russian Academy of Sciences, Bolshaya Gruzinskaya str. 10/1, 123242, Moscow, Russia }%

\author{Aleksey Khlyupin}%

\affiliation{%
 Moscow Institute of Physics and Technology, 9 Institutskiy per., Dolgoprudny, Moscow Region, 141701, Russian Federation}%

\author{Kirill  Gerke}
\affiliation{Schmidt Institute of Physics of the Earth of Russian Academy of Sciences, Bolshaya Gruzinskaya str. 10/1, 123242, Moscow, Russia}%

\date{\today}

\begin{abstract}
Precise characterization of three-dimensional heterogeneous media is indispensable in finding the relationships between structure and macroscopic physical properties (permeability, conductivity, and others). The most widely used experimental methods (electronic and optical microscopy) provide high-resolution bi-dimensional images of the samples of interest. However, 3D material inner microstructure registration is needed to apply numerous modeling tools. Numerous research areas search for cheap and robust methods to obtain "full" 3D information about the structure of the studied sample from its 2D cuts. In this work, we develop a dynamic phase-retrieval stochastic reconstruction algorithm that can create 3D replicas from 2D original images - DDTF. The DDTF is free of artifacts characteristic of previously proposed phase-retrieval techniques. While based on a two-point $S_2$ correlation function, any correlation function or other morphological metrics can be accounted for during the reconstruction, thus, paving the way to the hybridization of different reconstruction techniques. In this work, we use two-point probability and surface-surface functions for optimization.  To test DDTF, we performed reconstructions for three binary porous media samples of different genesis: sandstone, carbonate, and ceramic. Based on computed permeability and connectivity ($C_2$ and $L_2$ correlation functions), we have shown that the proposed technique in terms of accuracy is comparable to the classic simulated annealing-based reconstruction method but is computationally very effective. Our findings open the possibility of utilizing DDTF to produce fast or crude replicas further polished by other reconstruction techniques such as simulated annealing or process-based methods. Improving the quality of reconstructions based on phase-retrieval by adding additional metrics into the reconstruction procedure is possible for future work.
\end{abstract}

\maketitle


\section{\label{sec:level1}Introduction}

  Precise characterization of three-dimensional heterogeneous media is indispensable in finding the relationships between structure and macroscopic physical properties (permeability, conductivity, and other properties). \cite{adler1992porous, sahimi2003heterogeneous, torquato2002random}.The most widely used experimental methods (electronic and optical microscopy) provide high-resolution bi-dimensional images of the samples of interest. However, 3D material inner microstructure registration is needed to apply modeling tools. Obtaining quality 3D information is technically more complicated and requires expensive equipment and qualified operators. Thus, porous media reconstruction from thin sections \cite{adler1990flow, yeong1998reconstructing} is actual and has many petroleum and food engineering \cite{tahmasebi2012reconstruction, derossi2019mimicking}, material and soil sciences,\cite{jiao2014modeling, karsanina2015universal}, electrochemistry \cite{suzue2008micro}, and medical applications \cite{pellot19943d}. 
  Effective reconstruction procedures allow the creation of high-quality digital models for further analysis and calculation of physical properties. This is true for various multiphase materials, including porous media with a binary structure consisting of pores and solids. A versatile set of methods is available in the literature for stochastic reconstructions, including creating 3D replicas from 2D slices.
  Truncated Gaussian random fields \cite{hyman2014stochastic, roberts1997statistical} are a relatively computationally inexpensive approach but do not always allow the preservation of morphological properties of porous media. 
  Simulated annealing (SA) using correlation functions \cite{cule1999generating, ballani2015reconstruction} allows obtaining high-quality reconstructions, but it is a long-duration process even when accelerated (e.g., hierarchical simulated annealing) \cite{campaigne2012frozen, alexander2003hierarchical}). Note that hierarchical annealing schemes with correlation function rescaling were proposed recently to overcome this computational disadvantage \cite{karsanina2018hierarchical}. Multiple-point statistics (MPS) \cite{tahmasebi2012reconstruction, hajizadeh2011multiple, gravey2020quicksampling} reconstruction was shown to reproduce morphology quite well compared to other methods. 
  However, it is computationally expensive and may produce repetitive textures based on the number of points used for statistics. Quite surprisingly, some MPS methods are even slower and less accurate than SA based on two-point statistics \cite{lemmens2019nested}.  
  Deep learning approaches \cite{mosser2017reconstruction, feng2020end, coiffier20203d} are getting popular and show great promise, but even more computationally expensive than MPS and SA, with lower accuracies not balanced by massive training times. Process-based methods are very efficient and accurate, but their applicability is limited to granular porous media \cite{oren2002process, jin2003physics}.
  Phase-retrieval \cite{fullwood2008microstructure} is a fast algorithm that allows reconstructing volume from its two-point statistics \cite{torquato2002random} (or two-point probability correlation function). A method for approximating three-dimensional two-point statistics by a set of two-dimensional ones has been developed by Hasanabadi et al \cite{hasanabadi2016efficient}, and a three-dimensional image has been reconstructed from a bi-dimensional image. This approach has high performance and retains the basic microstructural properties of the media. However, the resulting image is rather noisy: artifacts like stripes and individual misplaced voxels appeared in the reconstructed sample. While this had only a minor impact on simulated electrical conductivity, other physical properties, especially permeability, are very sensitive to such noise \cite{gerke2015improving}. The originally described phase-retrieval method \cite{fullwood2008microstructure} used two-point statistics between each possible combination of pairs of pixels/voxels. This way, it was possible to reconstruct the images exactly \cite{chubb2000every}. However, in this case, the amount of two-point statistics is larger than the 2D or 3D image itself (even if we discard the parts redundant due to symmetry). Such reconstructions do not seem to have much practical application, especially considering the fact that two-point statistics that can be measured experimentally, for example, with the help of small-angle scattering \cite{debye1957scattering} or X-ray tomography \cite{li2018direct}, is limited to ensemble average over the whole volume or some direction. Ensemble statistics is a kind of compressed two-point statistics and can be effectively utilized to reduce the structural information about the object at hand \cite{gerke2015universal, havelka2016compression, karsanina2020compressing}. On the other hand, if the aim is to reconstruct 3D information from the 2D slice, using the complete information available from such a slice using full two-point statistics may be beneficial. In this work, we develop a phase-retrieval algorithm of three-dimensional sample reconstruction from its bidimensional cut-section free of the abovementioned disadvantages.\\
  The paper is organized as follows. We review the mathematics behind the phase-retrieval algorithm as well as the algorithm itself. Next, we describe rotation as a method for increasing the dimensionality of self-convolution using the fullest possible two-point statistics read from a 2D image. A new phase retrieval algorithm based on additional constraints in Fourier space and static and dynamic adjustments of three-dimensional self-convolution is presented. To demonstrate achieved improvements, we simulate single-phase (or saturated) flow on both original and reconstructed samples of three porous media samples of different genesis. We also evaluate the two-point cluster correlation function known to provide non-trivial connectivity information.
\section{\label{sec:MainPart}Methodology}
\subsection{Microstructure reconstruction from self-convolution}
\label{sec:1}
Microstructure $m^n_s$ is a binary multidimensional array where values 0 and 1 correspond to void and solid voxels, respectively. $(s_1, ..., s_N): s_1 \in (1, 2, ...,S_1), s_N \in (1, 2, ... S_N)$ is a vector of space coordinates while $n$ enumerates different components present in certain images, and $N$ is the dimension of space to which microstructure appertains.\\
Term correlation is commonly used in signal processing for convolution of the image with inversed self. However, this term in statistics has another meaning. Thus in this article, correlation in the sense of image processing is called convolution, while correlation is obtained as a result of either averaging convolutions by different directions in one image or by image ensemble averaging.
\\
Convolution of $m^n_s$ is defined the following way:
$$f^{nn'}_t = \frac{1}{S_1 \cdot ... \cdot S_n}\sum_{s_1 = 1}^{S_1}...\sum_{s_N = 1}^{S_N}m^n_s m^{n'}_{s+t},$$ where $S = (S_1, ..., S_N)$, $t = (t_1, ..., t_N) \in  R^n$. It contains sufficient data for retrieval of the initial binary image to within a translation and inversion. Fulwood et al.  establish the link between Fourier transform of microstructure and self-convolution so that the Gershberg-Saxton algorithm of image retrieval from magnitude of its Fourier transform becomes applicable.
\begin{multline}
M^n_k = DFT(m^n_s) = \\
\frac{1}{(S_1 \cdot ... \cdot S_n)}\sum_{s_1 = 1}^{S_1}...\sum_{s_N = 1}^{S_N}m^n_se^{2\pi itk/(S_1 \cdot ... \cdot S_n)},
\end{multline}
where $k = (k_1, ..., k_N) \in R^n$.
\begin{multline}
F^{nn}_k = \frac{1}{(S_1 \cdot ... \cdot S_n)}\sum_{s_1 = 1}^{S_1}...\sum_{s_N = 1}^{S_N}f^{nn}_se^{2\pi isk/(S_1 \cdot ... \cdot S_n)}\\
F^{nn}_k = \frac{1}{(S_1 \cdot ... \cdot S_n)}|M^n_k|^2.
\end{multline}
The resulting algorithm \ref{alg:GS} consists of main stages.
\begin{algorithm}[H]
\caption{Gershberg-Saxton algorithm}\label{alg:GS}
\begin{algorithmic}[1]
\STATE Calculation of microstructure's Fourier transform magnitude from self-convolution
$|M_k^1| = \sqrt{(S_1 \cdot ... \cdot S_n)F_k^{11}}$.
\STATE Making random binary noise initial guess for recovered microstructure  $(m^1_s)_0$.
\label{itm:start}
\REPEAT
\STATE Fourier transform of microstructure on current iteration.\\
$(M_k^1)_j = DFT((m_s^1)_j)$
\STATE Replacement its magnitude with the magnitude obtained from self-convolution. \\
$(M_k^{11})_j^{'} = |M^1_k| \cdot e^{i \cdot angle((M^1_k)_j)}$
\STATE Inverse Fourier transform of replacement result. \\
$(m_s^1)_j^{'} = IDFT((M_k^1)_j)$
\STATE Satisfying constraints in real space:
\begin{equation*}
  (m_s^1)_{j + 1} =
    \begin{cases}
      0 & \text{if $(m_s^1)_j^{'} \leq 0$}\\
      m^n_s & \text{if $0 \leq (m_s^1)_j^{'} \leq 1$}\\
      1 & \text{if $(m_s^1)_j^{'} \geq 0$}
    \end{cases}       
\end{equation*}
\STATE j++
\UNTIL{Residual $\delta = \sum_t\|f^{11}_t-IDFT((M_k^1)_j^* (M_k^1)_j)\| \leq \epsilon $ or maximum iteration number exceeds.}

\end{algorithmic}
\end{algorithm}
Using black padding around the input image before calculating its self-convolution and taking into account the position of padding during recovery results in a better convergence and reduction of translation uncertainty \cite{gaur2019sparsity}.
Nevertheless, in the case of approximated convolution reconstructed image with padding tends to have lower porosity in the slices adjacent to padding. Thus usage of padding is appropriate in cases when the role of boundary effects does not seriously influence macroscopic properties like porosity and permeability.

\subsection{The transition from 2d convolution to 3d convolution via rotation}
\begin{figure}
     \centering
         \centering
         \includegraphics[width=\columnwidth]{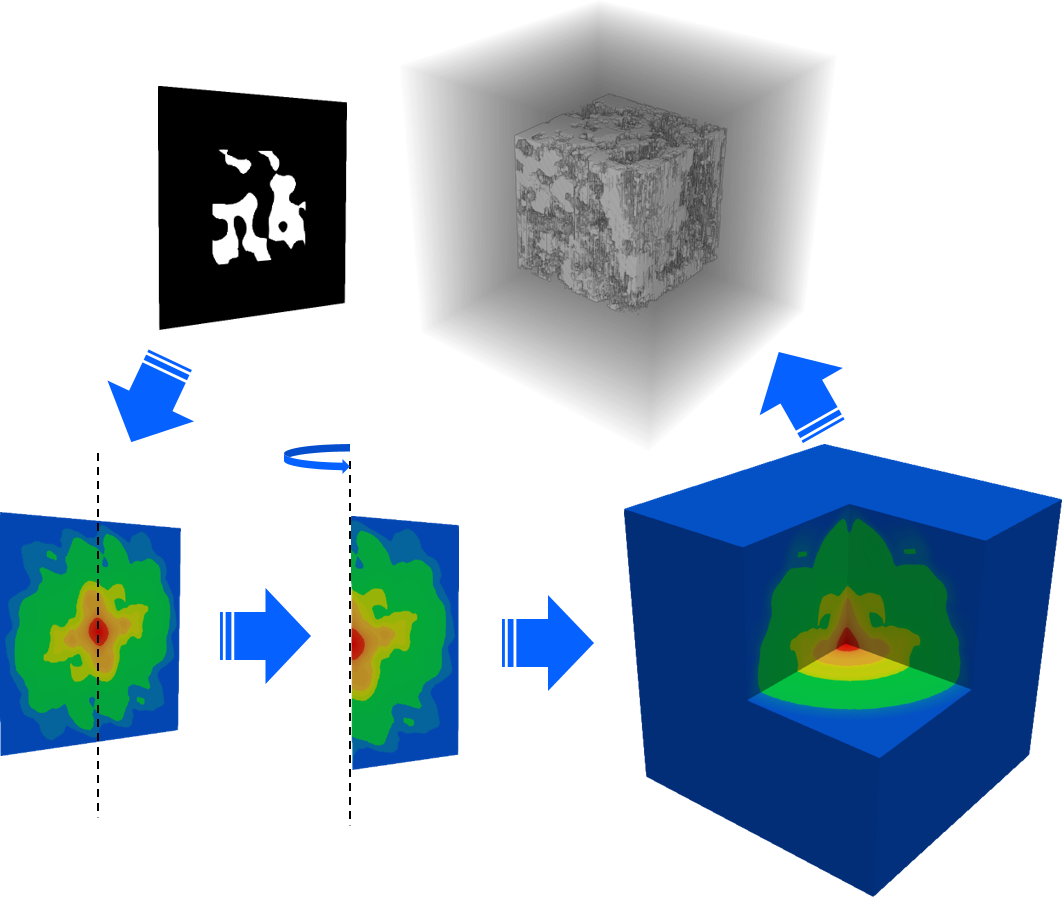}
         \caption{Reconstruction scheme}
         \label{fig:recscheme}
\end{figure}

\begin{figure}
     \centering
         \begin{subfigure}[t]{0.32\columnwidth}
                  \centering
         \includegraphics[width=\textwidth]{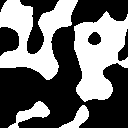}
         \caption{Input image}
         \end{subfigure}
         \hfill
        \begin{subfigure}[t]{0.32\columnwidth}
                 \centering
         \includegraphics[width=\textwidth]{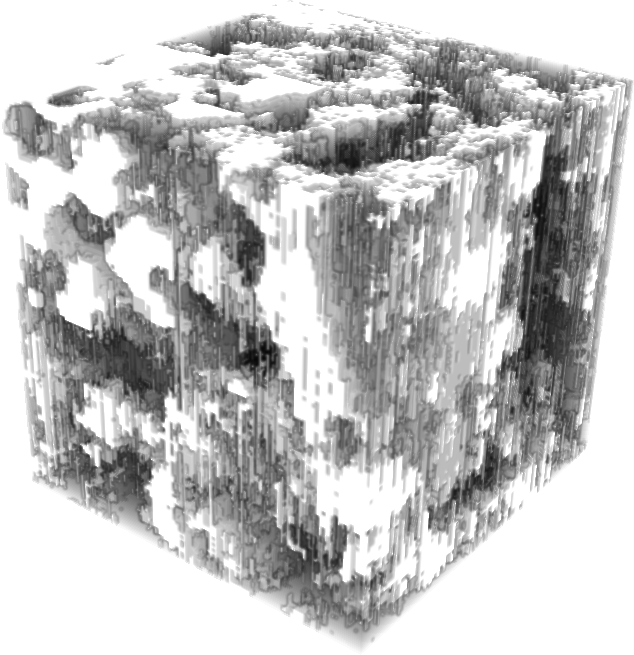}
         \caption{Reconstruction from self-convolution with preferential vertical direction}
         \end{subfigure}
        \hfill
        \begin{subfigure}[t]{0.32\columnwidth}
                 \centering
         \includegraphics[width=\textwidth]{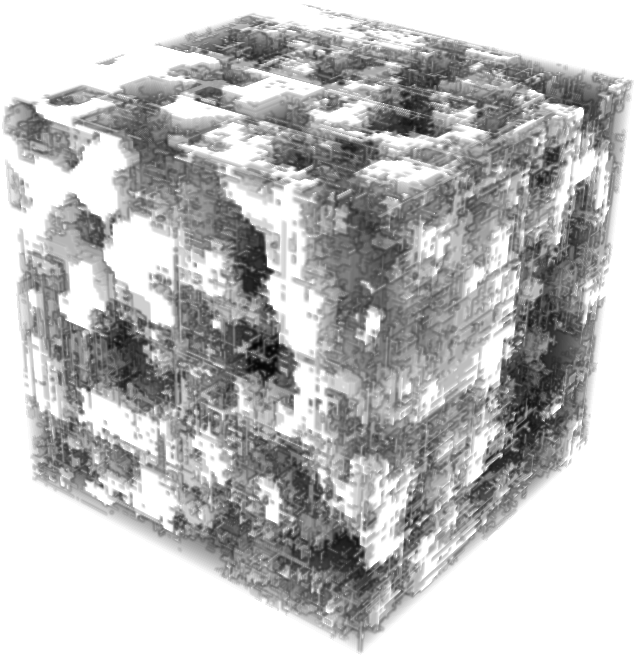}
          \caption{Reconstruction from mean of self-convolutions with preferential OX, OY, OZ directions}
         \end{subfigure}
         \caption{Influence if self-convolution anisotropy on reconstruction result}
         \label{fig:anis}
\end{figure}

One way to obtain a one-dimensional correlation for isotropic microstructure is by averaging its self-convolution in all directions. Self-convolution has axial symmetry. We propose rotation as a way to translate convolution from bi-dimensional to three-dimensional space.
Appropriate transition from 2d to 3d convolution is rather important, as it results in quality of reconstruction.
\\
In representative volume of porous media there is linear relationship between expected  values of 3d-convolution and 2d-convolution for equal shift values.
For binary image $m^n_s$ expectation of 2d-convolution for shift $(t, 0)$ is
\begin{multline}
E(f^{11}_{(t, 0)}) = \frac{1}{S_1\cdot S_2}E(\sum_{s_1 = 1}^{S_1}\sum_{s_2 = 1}^{S_2}m^n_s m^n_{s+(t, 0)}) =\\
= \frac{1}{S_1\cdot S_2} S_1 S_2 p = p,
\label{matec1}
\end{multline}
where $p$ is intersection porosity of original and shifted image.
For 3d-convolution for shift $(t, 0, 0)$ expectation is
\begin{multline}
E(f^{11}_{(t, 0, 0)}) = \frac{1}{S_1\cdot S_2 \cdot S_3}E(\sum_{s_1 = 1}^{S_1}\sum_{s_2 = 1}^{S_2}\sum_{s_3 = 1}^{S_3}m^n_s m^n_{s+(t, 0)}) =\\
=\frac{1}{S_1\cdot S_2 \cdot S_3} S_1 S_2 S_3 p = p = E(f^{11}_{(t, 0)})
\label{matec2}
\end{multline}\\
In case of adding black padding and doubling the linear size of an image $S' = 2S$ 
$$
    E(f^{11}_{(t, 0, 0)}) = \frac{p}{8} = \frac{1}{2} E(f^{11}_{(t, 0)})
$$
The approximation with properties mentioned above can be used as input for the phase-recovery algorithm.\\
Sample retrieved from this 3d-convolution has a preferential direction which is the convolution rotation axis. Rotation of 2d-convolution around ox, oy, oz axis and averaging the results lead to isotropy of reconstructed image (\ref{fig:anis}).
 \\
Three-dimensional convolution does not always correspond to existent microstructure, resulting in defects in a reconstructed image such as deformation and noise (figure \ref{fig:recscheme}).

\subsection{Frequency filtration}
\begin{figure}
     \centering
         \begin{subfigure}[t]{0.32\columnwidth}
                  \centering
         \includegraphics[width=\textwidth]{1592-.png}
         \caption{No filtration}
         \end{subfigure}
         \hfill
        \begin{subfigure}[t]{0.32\columnwidth}
                 \centering
         \includegraphics[width=\textwidth]{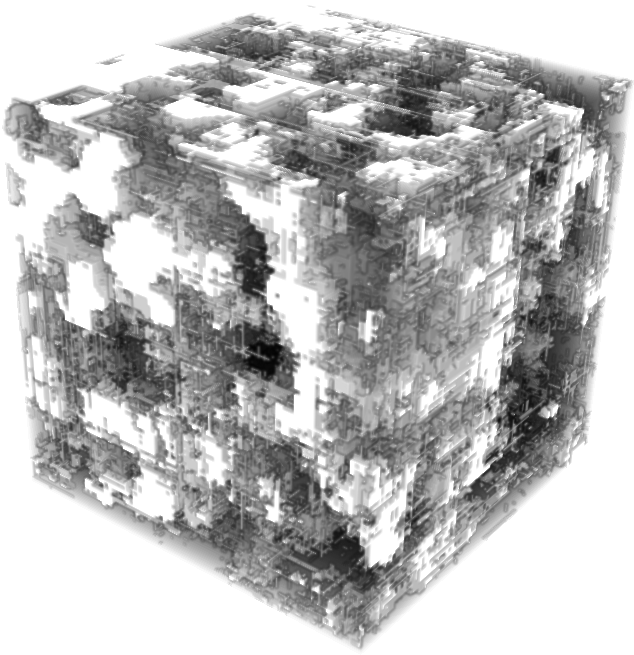}
         \caption{Width 0.5 px}
         \end{subfigure}
        \hfill
        \begin{subfigure}[t]{0.32\columnwidth}
                 \centering
         \includegraphics[width=\textwidth]{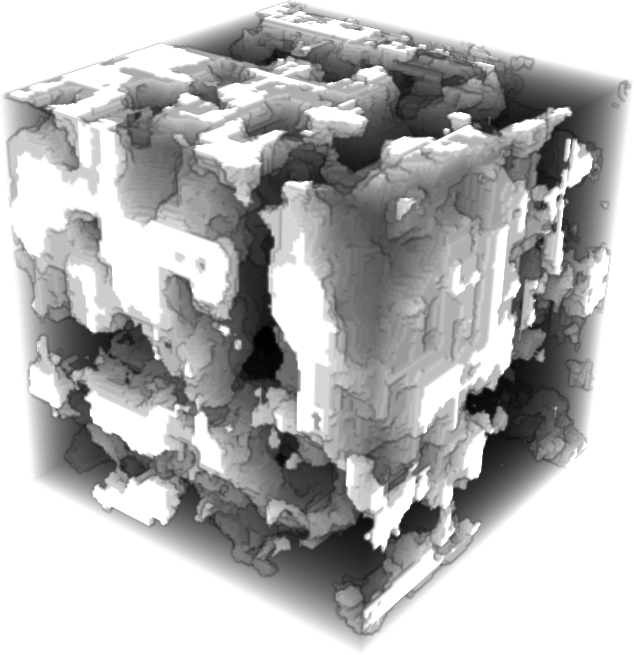}
          \caption{Width 1 px}
         \end{subfigure}
         \hfill
                 \raggedleft
        \begin{subfigure}[t]{0.32\columnwidth}
\raggedleft
         \includegraphics[width=\textwidth]{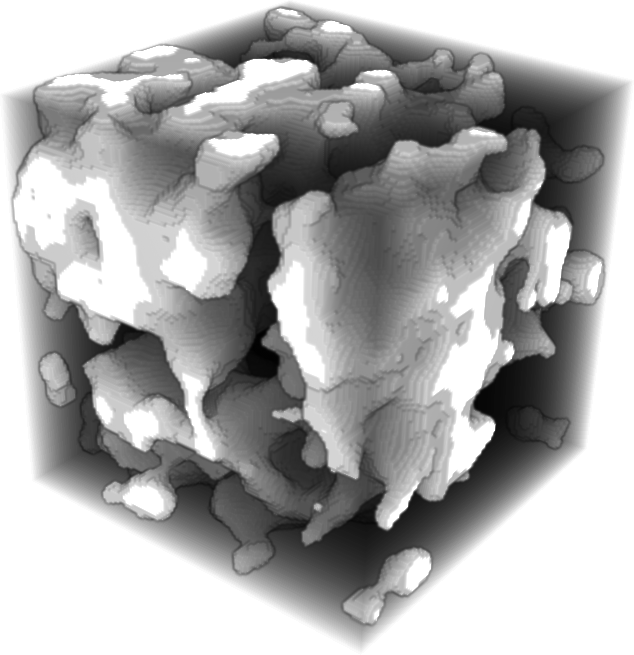}
          \caption{Width 1.5 px}
         \end{subfigure}
        \begin{subfigure}[t]{0.32\columnwidth}
                \raggedleft
         \includegraphics[width=\textwidth]{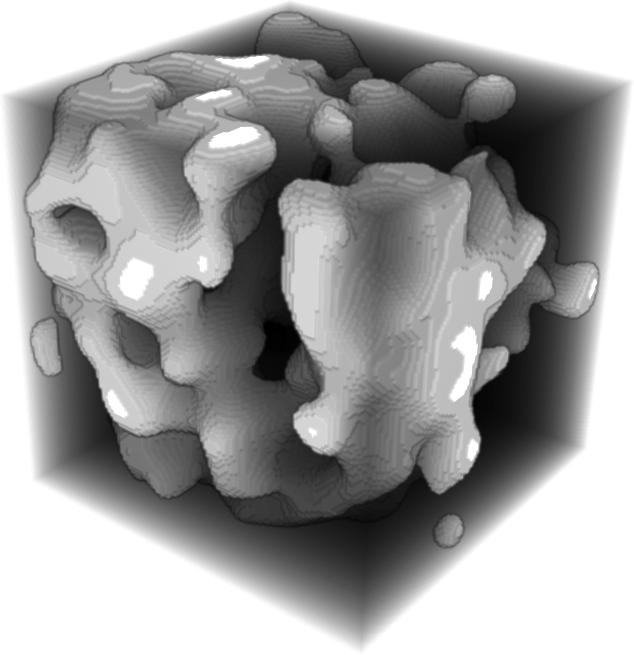}
          \caption{Width 2 px}
         \end{subfigure}
         \caption{Reconstruction with different Gaussian filters widths}
        \label{fig:filtr}
\end{figure}

Adding constraints in Fourier space such as frequency filtering is an efficient way of noise reduction.
Further in the article, samples are presented without padding for better presentation and visibility.
Though filter adjustment should be performed in each specific case, the width of the corresponding filter in real space should be no less than one voxel.\\
Figure \ref{fig:filtr} shows the reconstructed samples. The first image is obtained with no filtering. The following images correspond to space filters' widths of 0.5, 1, 1,5, and 2 voxels. Choice of appropriate mask width is the solution to the optimization problem for the cost function. For example, it is possible to minimize the difference between the correlation of the input image and the median correlation calculated of all slices. Stochastic optimization is warranted here because of a significant number of local minima. Thus modified algorithm \ref{alg:FreqGS} at step \ref{ff} also contains multiplication.
\begin{algorithm}[H]
\caption{Gershberg-Saxton algorithm with frequency filtration}\label{alg:FreqGS}
\begin{algorithmic}[1]
\STATE Calculation of microstructure Fourier transform magnitude from self-convolution 
$|M_k^1| = \sqrt{(S_1 \cdot ... \cdot S_n)F_k^{11}}$.
\STATE Making random binary noise the initial guess for recovered microstructure $(m^1_s)_0$.
\label{itm:start}
\REPEAT
\STATE Fourier transform of microstructure on current iteration.\\
$(M_k^1)_j = DFT((m_s^1)_j)$
\STATE Replacement its magnitude with the magnitude obtained from self-convolution. \\
$(M_k^{11})_j^{'} =|M^1_k| \cdot e^{i \cdot angle((M^1_k)_j)}$
\STATE Frequency filtration with Gaussian filter \label{ff}
\STATE Inverse Fourier transform of replacement result. \\
$(m_s^1)_j^{'} = IDFT((M_k^1)_j)$
\STATE Satisfying constraints in real space:
\begin{equation*}
  (m_s^1)_{j + 1} =
    \begin{cases}
      0 & \text{if $(m_s^1)_j^{'} \leq 0$}\\
      m^n_s & \text{if $0 \leq (m_s^1)_j^{'} \leq 1$}\\
      1 & \text{if $(m_s^1)_j^{'} \geq 0$}
    \end{cases}       
\end{equation*}
\STATE j++
\UNTIL{Maximum iteration number exceeds.}

\end{algorithmic}
\label{algf}
\end{algorithm}

\subsection{Dimensional effects}

\subsubsection{Examples of bidimensional and three-dimensional correlations}
Figure \ref{fig:dimef}  shows that transition functions from 2d-correlations to 3d-correlation for such objects as square, circular disc, and 2d-plurigaussian are non-linear and sufficiently different. We refer to such transition functions for correlations as "Dimensional Transition Functions" or DTF.
\subsubsection{Static dimensional transition functions}
\begin{figure}
     \centering
     \begin{subfigure}[t]{0.32\columnwidth}
         \centering
         \includegraphics[width=\textwidth]{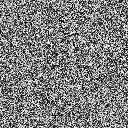}
         \caption{Random uniform noise $U$}
     \end{subfigure}
     \hfill
     \centering
     \begin{subfigure}[t]{0.32\columnwidth}
         \centering
         \includegraphics[width=\textwidth]{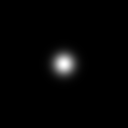}
         \caption{Gaussian kernel $K$}
     \end{subfigure}
     \hfill
          \centering
     \begin{subfigure}[t]{0.32\columnwidth}
         \centering
         \includegraphics[width=\textwidth]{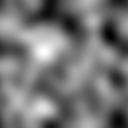}
         \caption{Convolution of noise with kernel $T=K*U$}
     \end{subfigure}
     \hfill
     \centering
     \begin{subfigure}[t]{0.32\columnwidth}
         \centering
         \includegraphics[width=\textwidth]{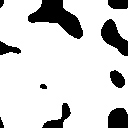}
         \caption{$T$ thresholded by 0.3 from maximal value}
     \end{subfigure}
     \hfill
     \centering
     \begin{subfigure}[t]{0.32\columnwidth}
         \centering
         \includegraphics[width=\textwidth]{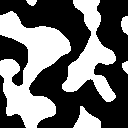}
         \caption{$T$ thresholded by 0.5 from maximal value}
     \end{subfigure}
     \hfill
          \centering
     \begin{subfigure}[t]{0.32\columnwidth}
         \centering
         \includegraphics[width=\textwidth]{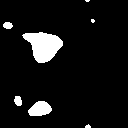}
         \caption{$T$ thresholded by 0.7 from maximal value}
     \end{subfigure}
     \hfill
     \caption{Generating truncated Gaussian random fields by convolution of uniform random noise with Gaussian kernel}
     \label{fig:trf}
\end{figure}

Thus approximated 3d-convolution should be adjusted later on account of microstructure morphology. \\
Plurigaussian modeling of microstructures is one of the widely used methods \cite{hyman2014stochastic} for microstructure modeling. Random topography field $T$ is obtained by convolving kernel $k(x)$ with  $u(x) \in U$ - random field with uniform probability density distribution.\\
$$
T(x) = \int_{R^n}k(x-y)u(y)dy
$$
The convolution of the Gaussian kernel with the random field is shown in figure \ref{fig:trf}.
\begin{figure}
     \centering
     \begin{subfigure}[t]{0.32\columnwidth}
         \centering
         \includegraphics[width=\textwidth]{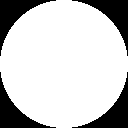}
         \caption{Solid sphere slice}
     \end{subfigure}
     \hfill
     \centering
     \begin{subfigure}[t]{0.32\columnwidth}
         \centering
         \includegraphics[width=\textwidth]{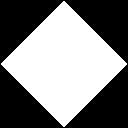}
         \caption{Thetraedron slice}
     \end{subfigure}
     \hfill
     \centering
     \begin{subfigure}[t]{0.32\columnwidth}
         \centering
         \includegraphics[width=\textwidth]{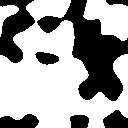}
         \caption{Plurigaussian slice}
     \end{subfigure}
     \hfill
     \\
    \centering
     \begin{subfigure}[t]{\columnwidth}
         \centering
         \includegraphics[width=\textwidth]{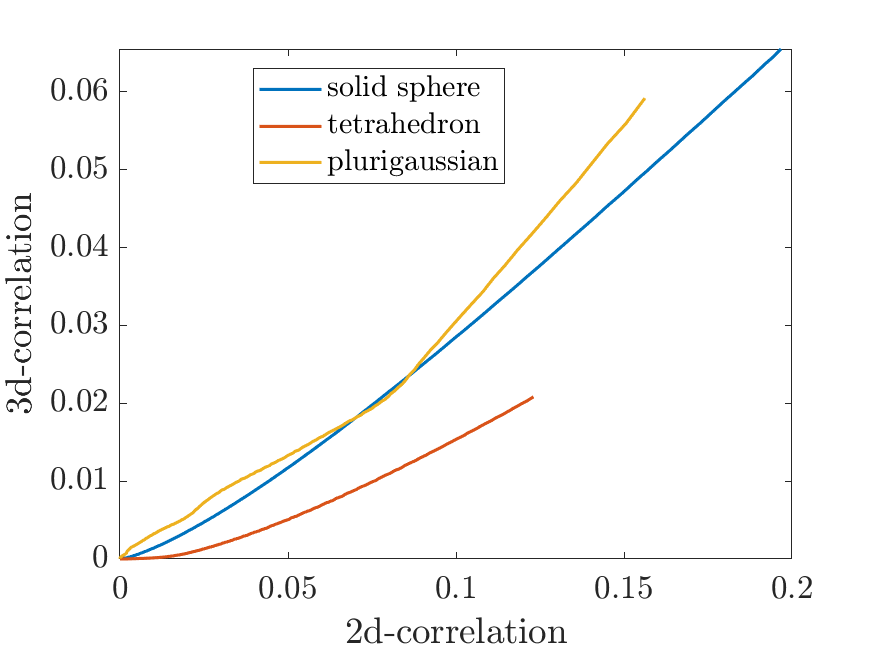}
         \caption{Dimensional transition functions for solid sphere, thetraedron, plurigaussian}
     \end{subfigure}
        \caption{Dimensional transition functions for solid sphere, thetraedron, plurigaussian}
        \label{fig:dimef}
\end{figure}

We implemented an algorithm \ref{alg:SDTF} to generate an approximated microstructure in which 2d-slice corresponds with the input image. Its realization is described thereunder.
\begin{algorithm}[H]
\caption{SDTF preprocessing algorithm}\label{alg:SDTF}
\begin{algorithmic}[1]
\STATE Calculate the self-convolution of a 2d-image. 
\STATE Average self-convolution by all directions to obtain 1D-correlation. 
\STATE Translate 1D-correlation to 3d-space by rotation. 
\STATE Generate a 3d-random field with uniform distribution.  
\STATE Convolve kernel obtained by rotation of self-convolution and random field with uniform distribution. 
\STATE Truncate the resulting plurigaussian so that the porosity of the middle slice of the resulting microstructure is the closest to the porosity of the input image.
\STATE Circularly shift truncated plurigaussian for alignment with input slices using a maximum of cross-convolution.
\STATE Apply morphological erosion or dilation to minimize differences between correlation functions of the input slice and the particular slice of truncated plurigaussian. 
\STATE Calculate correlation function from radial shift for certain 2D-slice of plurigaussian approximation. 
\STATE Calculate correlation function from radial shift for the whole plurigaussian approximation.
\STATE Plot set of points. Each point corresponds to a particular radial shift where the x coordinate is the value for 2d-correlation, and the y-value is 3d-correlation. 
\STATE Approximate the dependency by power-law to apply it to 3d-autoconvolution obtained by rotation from bidimensional one
\end{algorithmic}
\end{algorithm}

Such plurigaussian microstructure is used only for building dimensional transition function in each specific case. \\
A synthetic case like circular disk (figure \ref{fig:dimef}) is an example when appropriate convolution transform is essential. Using traditional scaling, we obtain significantly worse results while plurigaussian approximation for circular disk leads to the reconstruction of an object similar to a solid sphere.\\
In the case of small-scale microstructure, DTF-correction does not play a significant role because the transition function is almost linear. Properties of the resulting sample differ in relation to noise realization because the kernel is convolved with random noise. Thus it is necessary to choose the best realization according to quality metrics such as correlation function. However, in the case of images whose characteristic scale is comparable with its size, plurigaussian approximation does not always give a reasonable guess for dimensional transitional coefficients.
\subsubsection{The dynamic dimensional transition function}
\begin{figure}
     \centering
     \begin{subfigure}[t]{0.32\columnwidth}
         \centering
         \includegraphics[width=\textwidth]{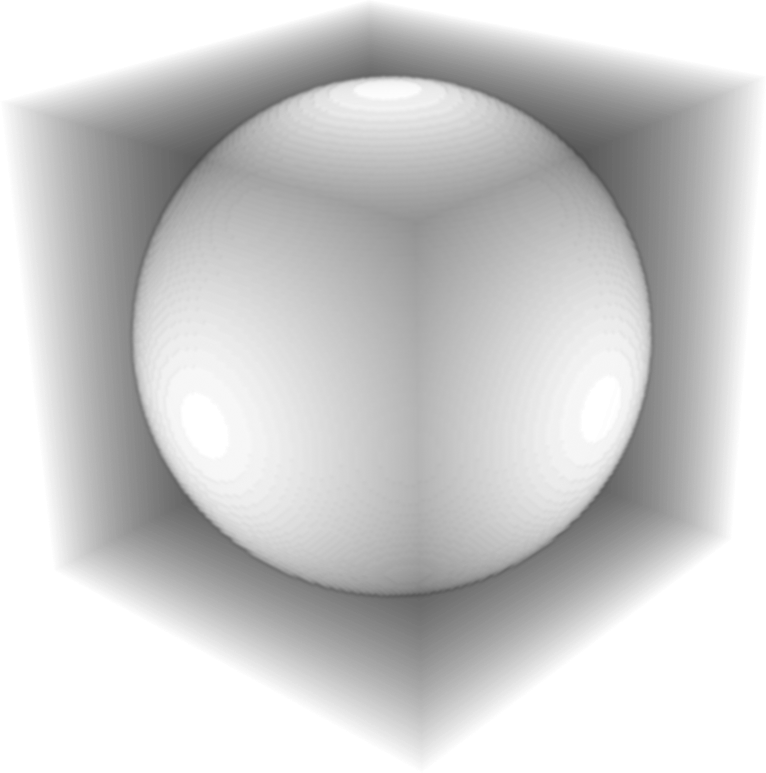}
         \caption{Original image}
         
     \end{subfigure}
     \hfill
     \centering
     \begin{subfigure}[t]{0.32\columnwidth}
         \centering
         \includegraphics[width=\textwidth]{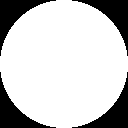}
         \caption{Input image}
         
     \end{subfigure}
     \hfill
     \centering
     \begin{subfigure}[t]{0.32\columnwidth}
         \centering
         \includegraphics[width=\textwidth]{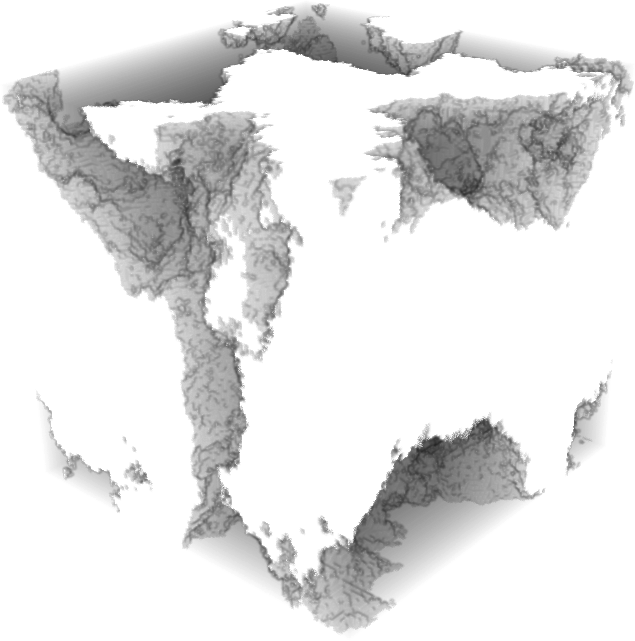}
         \caption{Plurigaussian approximation}
         
     \end{subfigure}
     \hfill
     \\
     \centering
     \begin{subfigure}[t]{0.32\columnwidth}
         \centering
         \includegraphics[width=\textwidth]{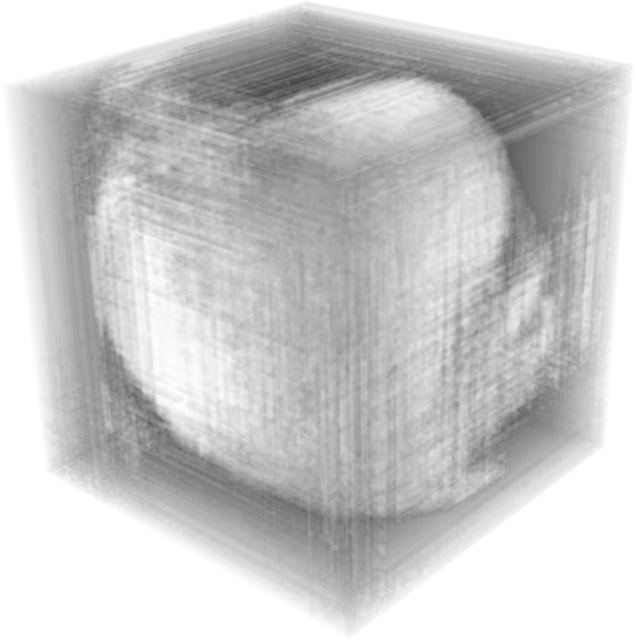}
         \caption{Static linear reconstruction}
         
     \end{subfigure}
     \hfill
     \begin{subfigure}[t]{0.32\columnwidth}
         \centering
         \includegraphics[width=\textwidth]{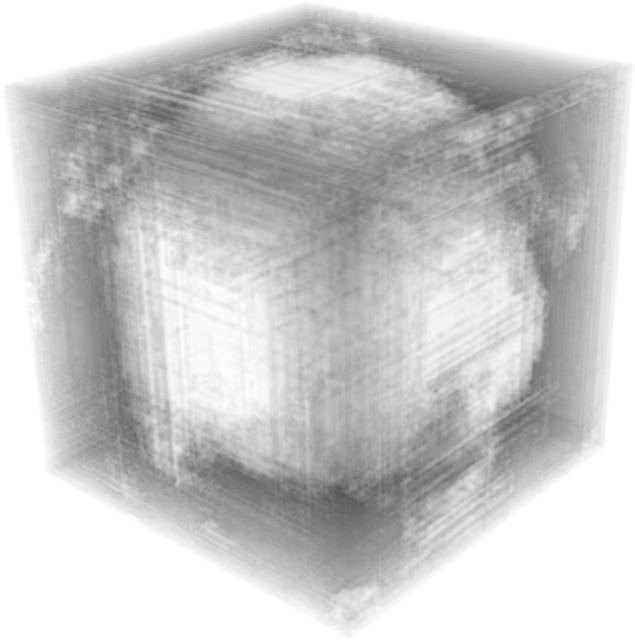}
         \caption{Static plurigaussain reconstruction}
     \end{subfigure}
     \hfill
     \begin{subfigure}[t]{0.32\columnwidth}
         \centering
         \includegraphics[width=\textwidth]{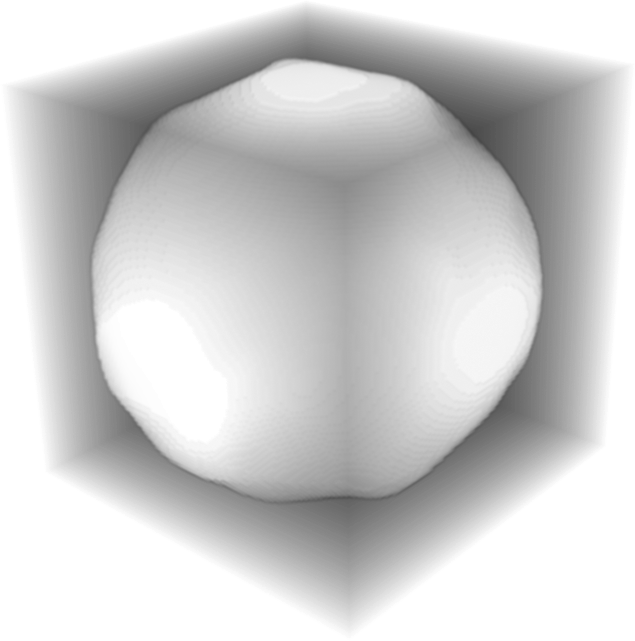}
         \caption{Dynamic reconstruction}
     \end{subfigure}
        \caption{Comparison of different reconstruction methods}
        \label{fig:compSDTF_DDTF}
\end{figure}

\begin{figure}
     \centering
     \begin{subfigure}[t]{0.32\columnwidth}
         \centering
         \includegraphics[width=\textwidth]{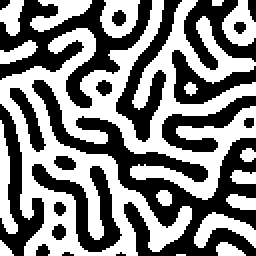}
         \caption{Turing pattern}
     \end{subfigure}
     \hfill
          \begin{subfigure}[t]{0.32\columnwidth}
         \centering
         \includegraphics[width=\textwidth]{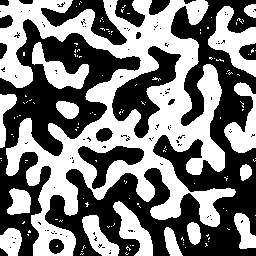}
         \caption{Slice of 3D-reconstruction}
         
     \end{subfigure}
     \hfill
          \begin{subfigure}[t]{0.32\columnwidth}
         \centering
         \includegraphics[width=\textwidth]{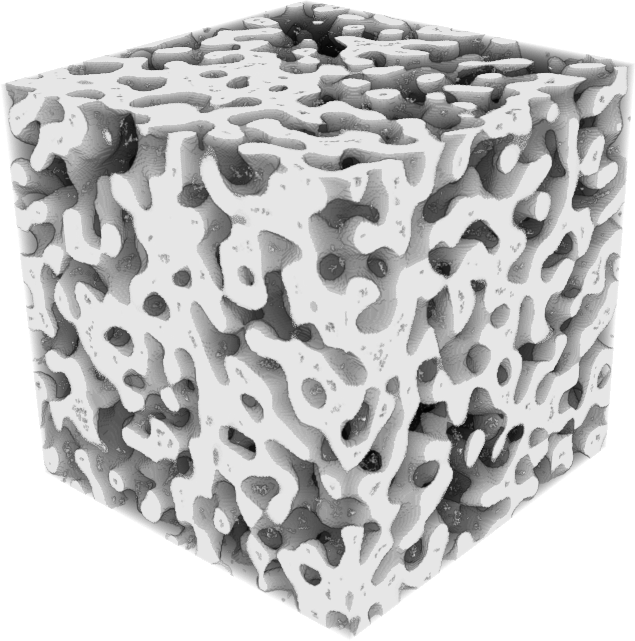}
         \caption{3D-reconstruction}
     \end{subfigure}
              \hfill
      \begin{subfigure}[t]{\columnwidth}
         \centering
         \includegraphics[width=\textwidth]{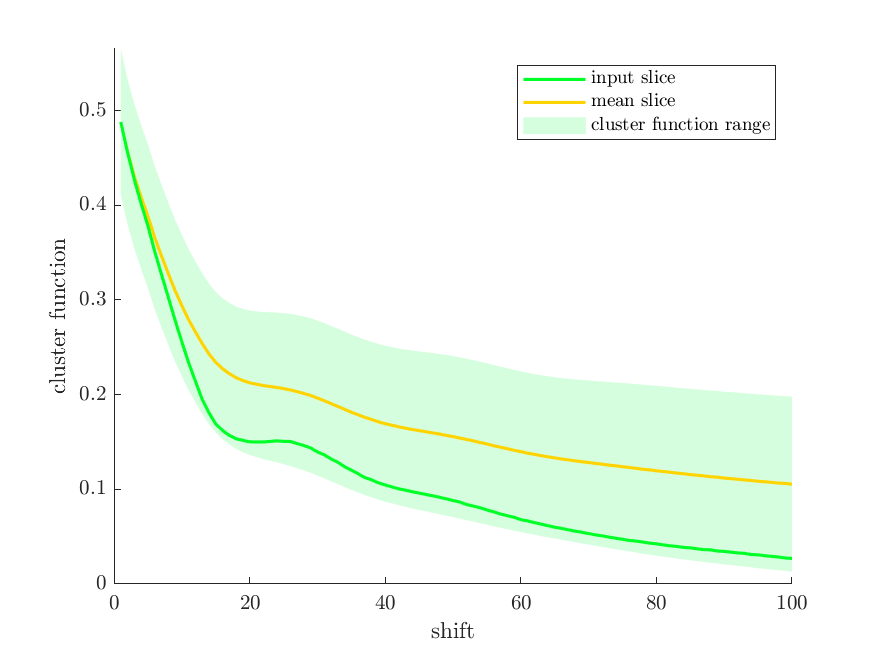}
         \caption{Cluster functions of input image and slices of reconstruction}
         
     \end{subfigure}
        \caption{Turing pattern reconstruction}
        \label{fig:turing}
\end{figure}

Quality for building both convolution and DTF influences reconstruction results. The solution is on-the-fly updating of the convolution in such a way that allows adjusting parameters of convolution during phase-retrieval algorithm execution. For example, parameters can be chosen to minimize difference according to a particular metric between a specific slice or slices of reconstructed and input image. The choice of metric is a single issue and depends on the integral characteristic of the 3d-image (\cite{torquato2002random}).\\
Approximated 3d-autoconvolution has radial and angular parts. The radial part is relatively smooth in all points except $(0, 0, 0)$, while the angular one somewhat oscillates around zero field with magnitude decreasing with the growth of distance from autocorrelation center $(0, 0, 0)$. Coefficients in the linear combination of these two parts in the resulting approximation for autoconvolution are also optimization parameters. Thus modified algorithm \ref{alg:DDTF} is the following.
\begin{algorithm}[H]
\caption{DDTF reconstruction algorithm}\label{alg:DDTF}
\begin{algorithmic}[1]
\STATE Calculation of microstructure Fourier transform magnitude from self-convolution, \\
$|M_k^1| = a*\sqrt{(S_1 \cdot ... \cdot S_n)F_k^{11}}^b$ \\
where a and b are dimensional transition coefficients.
\STATE $|M_k^1|_{lc} = rad(|M_k^1|) + c * ang(|M_k^1|)$, where $rad$ and $ang$ are radial and angular parts respectively, and $c$ is optimization coefficient.
\STATE Making random binary noise initial guess for recovered microstructure $(m^1_s)_0$.
\label{itm:start}
\REPEAT
\STATE Fourier transform of microstructure on current iteration.\\
$(M_k^1)_j = DFT((m_s^1)_j)$
\STATE Replacement of its magnitude with the magnitude obtained from self-convolution. \\
$(M_k^{11})_j^{'} =|M^1_k|_{lc} \cdot e^{i \cdot angle((M^1_k)_j)}$
\STATE Frequency filtration with Gaussian filter which width is optimization parameter $d$.
\STATE Inverse Fourier transform of replacement result. \\
$(m_s^1)_j^{'} = IDFT((M_k^1)_j)$\\
\STATE Calculating cost objective $\delta_{j}$ as the difference between the correlation function of the input image and the correlation function of a particular slice of reconstruction.\\
\STATE
\eIf{$\delta_{j}$ is $min(\delta(1), ..\delta(j))$}{
   Satisfying constraints in real space:
   \begin{equation*}
      (m_s^1)_{j + 1} =
        \begin{cases}
          0 & \text{if $(m_s^1)_j^{'} \leq 0$}\\
          (m_s^1)_j^{'} & \text{if $0 \leq (m_s^1)_j^{'} \leq 1$}\\
          1 & \text{if $(m_s^1)_j^{'} \geq 0$}
        \end{cases}       
    \end{equation*}
    }{
         \begin{equation*}
      (m_s^1)_{j + 1} =  (m_s^1)_{j}
    \end{equation*}
    }
\STATE Updating parameters a, b, c, d according to the optimization algorithm.
\STATE j++
\UNTIL{Maximum iteration number exceeds.}

\end{algorithmic}
\label{algf}
\end{algorithm}
For example, the correlation function shows good performance for the solid sphere mentioned above, even in the case of the wrong guess of SDTF.
Even in case of inappropriate guess for static DTF convolution can be successfully modified for satisfying constraints for 2d-slices (figure \ref{fig:compSDTF_DDTF}). There is no need to optimize parameters $a$ and $b$ due to the linear transition function for porous media. However, angular coefficient c and filter width d still need to be determined. In this case, the algorithm is the following.
\begin{algorithm}[H]
\caption{DDTF reconstruction for porous media}\label{alg:DDTF-Porous}
\begin{algorithmic}[1]
\STATE Calculation of microstructure Fourier transform magnitude from self-convolution, \\
$|M_k^1| = \sqrt{(S_1 \cdot ... \cdot S_n)F_k^{11}}$ \\
where a and b are dimensional transition coefficients.
\STATE $|M_k^1|_{lc} = rad(|M_k^1|) + c * ang(|M_k^1|)$, where $rad$ and $ang$ are radial and angular parts respectively, and $c$ is optimization coefficient.
\STATE Making random binary noise initial guess for recovered microstructure $(m^1_s)_0$.
\label{itm:start}
\REPEAT
\STATE Fourier transform of microstructure on current iteration.\\
$(M_k^1)_j = DFT((m_s^1)_j)$
\STATE Replacement its magnitude with the magnitude obtained from self-convolution. \\
$(M_k^{11})_j^{'} =|M^1_k|_{lc} \cdot e^{i \cdot angle((M^1_k)_j)}$
\STATE Frequency filtration with Gaussian filter which width is optimization parameter d. 
\STATE Inverse Fourier transform of replacement result. \\
$(m_s^1)_j^{'} = IDFT((M_k^1)_j)$\\
\STATE Thresholding $(m_s^1)_j^{'}$ to conserve porosity of input image.\\
$(m_s^1)_j^{''} = (m_s^1)_j^{'} > th$\\
$E((m_s^1)_j^{''}) = E((m_s^1)_{2d})$\\
where $th$ is thresholod and $(m_s^1)_{2d}$ is input image.
\STATE Calculating cost objective $\delta_j$ as negative likelihood function (appendix \ref{eq:lkhd}) for surface correlation function of the input image and all the slices of reconstruction.\\
\STATE Updating parameters c, d according to the optimization algorithm. 
\STATE j++
\UNTIL{Maximum iteration number exceeds.}

\end{algorithmic}
\label{algf}
\end{algorithm}
While Algorithms \ref{alg:DDTF-Porous} and \ref{alg:DDTF} look very similar, they possess some principal differences, the initial guess being the most pronounced one. In Algorithm 5, it is always the same random noise. On the other hand, in Algorithm 4, the random noise is used as the first guess, and each next iteration uses the best previous results as input. Such an implementation allows combining optimization steps consisting of phase recovery with different parameters to equivalent phase-recovery with a more significant number of iterations. The influence of this approach is significant in cases when the number of phase recovery iterations per one optimization step is not enough to reconstruct microstructure from random noise. This seems unnecessary for general porous media (unlike the ball reconstruction problem) due to the less localized distribution of voxels resulting in less iteration number necessary for the convergence. Note that a simple increase in iterations number per optimization step would resolve this problem but is expected to increase overall iterations.\\
Reconstruction of Turing pattern (which is a complex structure often used for algorithm attestation) by algorithm \ref{alg:DDTF-Porous} leads to the following results (figure \ref{fig:turing}). It can be seen that the slice of the reconstruction, as well as the original, contains both isolated solid and void areas of similar size, but the thickness of patterns is rarely conserved.

\subsubsection{The classical simulated annealing reconstruction technique}%
In order to benchmark the new methodology developed in this work, we compare the reconstruction results against a classical reconstruction based on correlation functions. To make both techniques equal, we applied modified Yeong-Torquato reconstruction  \cite{yeong1998reconstructing} and utilized only the two-point probability function $S_2(r)$ \cite{torquato2002random} describing the probability that two points separated by a vector displacement $r(x_1,x_2)$ between $x_1$ and $x_2$ lie in the pore phase. We calculate $S_2(r)$ functions in two orthogonal and two diagonal directions, thus, giving four independent CFs for each input 2D image, which are then used separately during reconstruction \cite{gerke2015improving}, which are averaged to estimate the other 4 CFs for the unknown third dimension. For any set of correlation functions considered in the Yeong-Torquato technique, matching correlation functions of a given realization with a target CFs set is based on pixel permutations. If a set of two-point correlation functions used in reconstruction is provided in the form of $f_2^\alpha(r)$, where $\alpha$ is a type of CF (different directions in our case) and $r$ is a segment of varying length, the difference between two realizations of the structure can be expressed as the sum of squared differences \cite{yeong1998reconstructing, gerke2015improving}:
\begin{equation}
    E = \sum_\alpha \omega_\alpha \sum_r[f_2^\alpha(r) - \hat{f}_2^\alpha(r)]^2
    \label{eq:cor}
\end{equation}
where $f_2^\alpha(r)$ and $\hat{f}_2^\alpha(r)$ are the values of the correlation function sets for two realizations (where the former represents a reference set while the latter represents current reconstruction state), $w_\alpha$ are weighting parameters chosen based on the input of each CF into energy E for disordered structure \cite{gerke2015improving}. The energy E in eq. \ref{eq:cor} is minimized by the simulated annealing optimization. The Metropolis algorithm is used to determine the probability of accepting any permutation p:
\begin{equation}
    p(E_{old} \rightarrow E_{new}) = \begin{cases}
    1, \Delta E < 0\\
    exp(-\frac{\Delta E}{T}), \Delta E \geqslant 0
    \end{cases}
    \label{eq:ann}
\end{equation}
where $T$ is the temperature of the system, and
\begin{equation}
    \Delta E = E_{new} - E_{old}
    \label{eq:dt}
\end{equation}

At initialization, the temperature is chosen so that the probability p for $\Delta E \geqslant 0$ equals 0.5 \cite{yeong1998reconstructing}. The following cooling schedule based on geometrical progression is used:
\begin{equation}
    T(k) = T(k - 1)\lambda
    \label{eq:tk}
\end{equation}
where $k$ is the time step and $\lambda$ is a parameter smaller than but close to unity ($\lambda$=0.999999 for all reconstructions presented here). An optimized Capek pixel permutation approach based on interface choices \cite{vcapek2009stochastic, vesely2015prediction} was used. Periodic boundary conditions were applied for CFs evaluation. The reconstruction procedure was terminated after $10^6$ consecutive unsuccessful permutations.
\subsection{Flow simulation}
Fluid flow in pore space is described by the Navier-Stokes equation (\ref{eq:NS}) \cite{batchelor1989fluid}.
\begin{equation}
 \begin{cases}
  \frac{\partial \textbf{v}}{\partial t} + (\textbf{v}\nabla)\textbf{v} - \frac{\mu}{\rho}\Delta\textbf{v}+\frac{\nabla p}{\rho}=0\\
  div \, \textbf{v} = 0
 \end{cases}
  \label{eq:NS}
\end{equation}
where \textbf{v}= $(v_x, v_y, v_z)$, $\mu$, $\rho$ and $p$ are velocity field, viscosity, density and pressure field respectively.\\
In the case of small Reynolds numbers $Re = \frac{\rho v l}{\mu} << 1$, which is typical for flow in porous media, equation \ref{eq:NS} can be transformed into the form (\ref{eq:NS2})
\begin{equation}
 \begin{cases}
  \rho \frac{\partial \textbf{v}}{\partial t} + \mu\Delta\textbf{v}+\nabla p=0 \, in\, \Omega\\
  div \, \textbf{v} = 0 \\
 \textbf{v}(\textbf{x},t) = 0\, on\, \partial\Omega\\
 \textbf{v}(\textbf{x},0) = 0
 \end{cases}
  \label{eq:NS2}
\end{equation}
where domain $\Omega$ is pore space and $\partial \Omega$ is its boundary.
Permeability $K$ can be determined form Darcy's law (\ref{Darcy})
\begin{equation}
K = \frac{\mu L Q}{\Delta p S} 
 \label{Darcy}
\end{equation}
where $\mu$ is viscosity, $L$ is the distance for which a pressure difference $\Delta p$ is applied, and $Q$ is flow rate through the cross-sectional area $S$. The set of equations (\ref{eq:NS2}) has been solved FDMSS solver with 4th order spatial accuracy scheme \cite{gerke2018finite}.  To ensure good convergence in all modelling cases either $3.5×10^3$ iterations were performed or we reached error criterion  $< 0.05$ which is based on imbalance for both continuity and motion parts of equation (\ref{eq:NS2}).
\subsection{Samples for reconstruction and comparison between techniques}
To verify the newly proposed modified phase-retrieval algorithm and compare it against classical technique, we chose three 3D porous media images of different genesis: artificial ceramic \cite{gerke2015studying}, sandstone and carbonate rocks \cite{gerke2020improving}. The choice was motivated by a wide range of porosities within these samples and their relative homogeneity and isotropy. Such properties are essential, as we assume the structures to statistically homogeneous so that we can consider r as a scalar distance between pixels while computing correlation functions and isotropy is needed to evaluate the unknown 3rd dimension to perform 2D into 3D reconstruction. We chose a single 2D slice with the porosity value closest to the porosity of the whole original 3D image as input data to both methods. Both simulated annealing and phase retrieval reconstructions are performed with periodic boundary conditions. To compare the reconstruction against the original 3D image, we compute error based on flow simulations using the following definition:

\begin{equation}
    K_{error} = \frac{K_{reconstruction}}{K_{original}} - 1
\end{equation}
where $K_{reconstruction}$ is the permeability of the stochastic reconstruction at hand, and $K_{original}$ is the simulated permeability based on the original 3D XCT image.
\section{Results and Discussions}
\begin{figure}
     \centering
     \begin{subfigure}[t]{0.32\columnwidth}
         \centering
         \includegraphics[width=\textwidth]{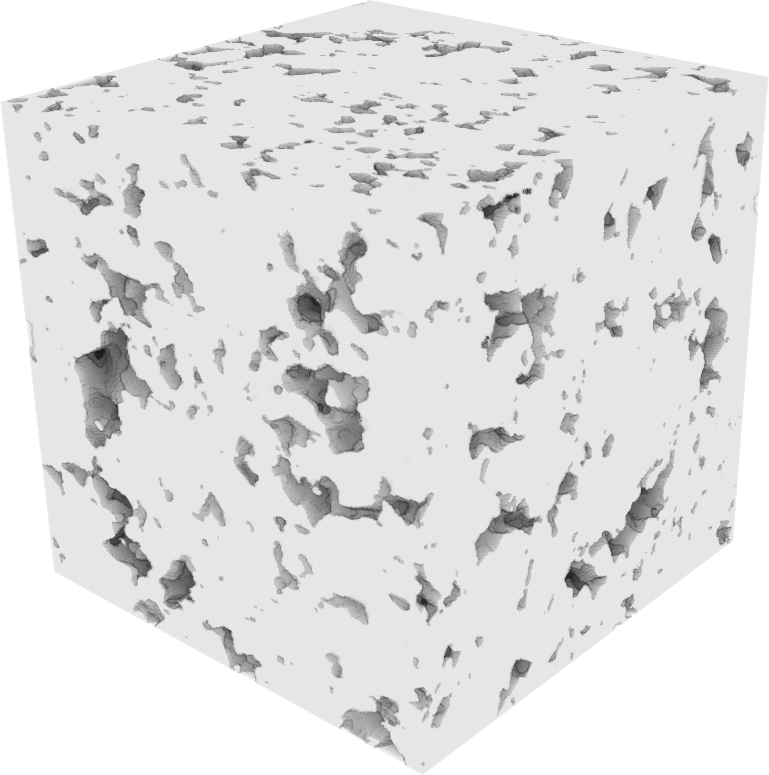}
         \caption{Carbonate original}
         
     \end{subfigure}
     \hfill
     \centering
     \begin{subfigure}[t]{0.32\columnwidth}
         \centering
         \includegraphics[width=\textwidth]{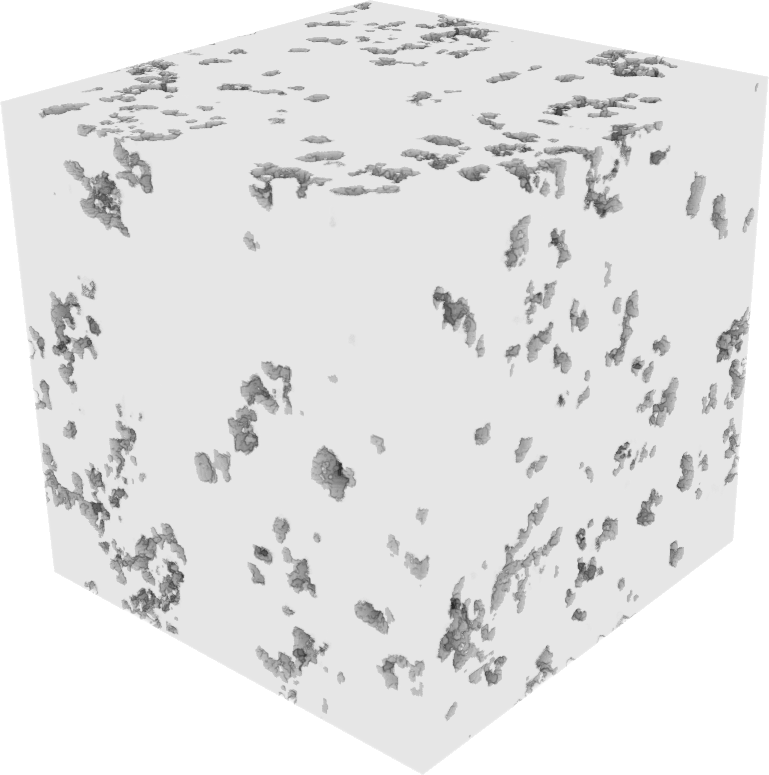}
         \caption{Carbonate SA reconstruction}
         
     \end{subfigure}
     \hfill
     \begin{subfigure}[t]{0.32\columnwidth}
         \centering
         \includegraphics[width=\textwidth]{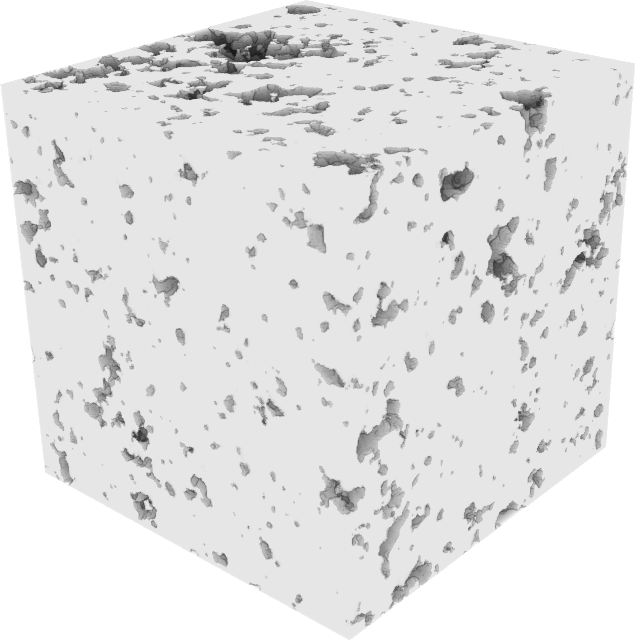}
         \caption{Carbonate PR reconstruction}
         
     \end{subfigure}
     \hfill
     \\
          \centering
     \begin{subfigure}[t]{0.32\columnwidth}
         \centering
         \includegraphics[width=\textwidth]{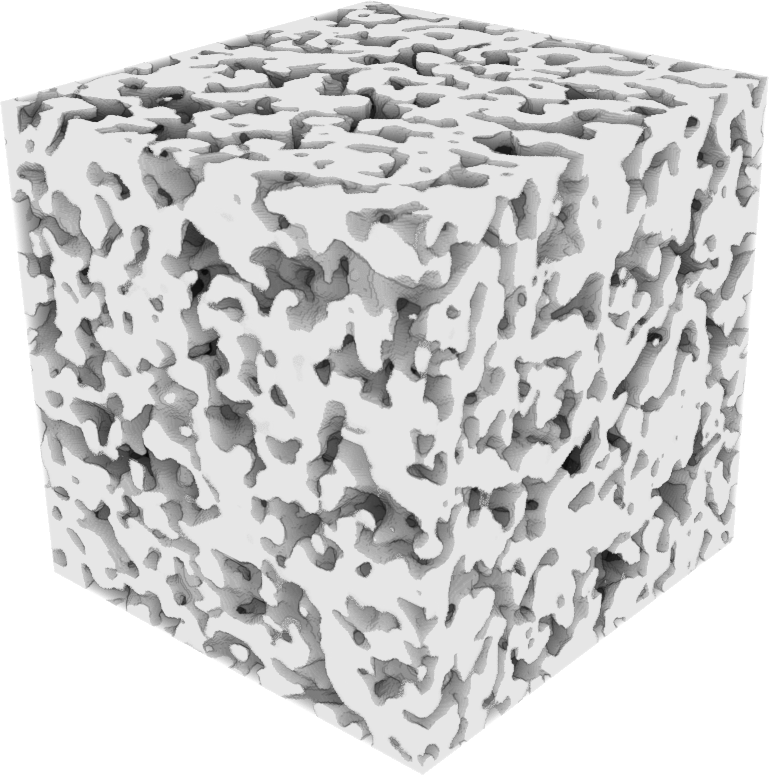}
         \caption{Ceramics original}
      
     \end{subfigure}
     \hfill
     \centering
     \begin{subfigure}[t]{0.32\columnwidth}
         \centering
         \includegraphics[width=\textwidth]{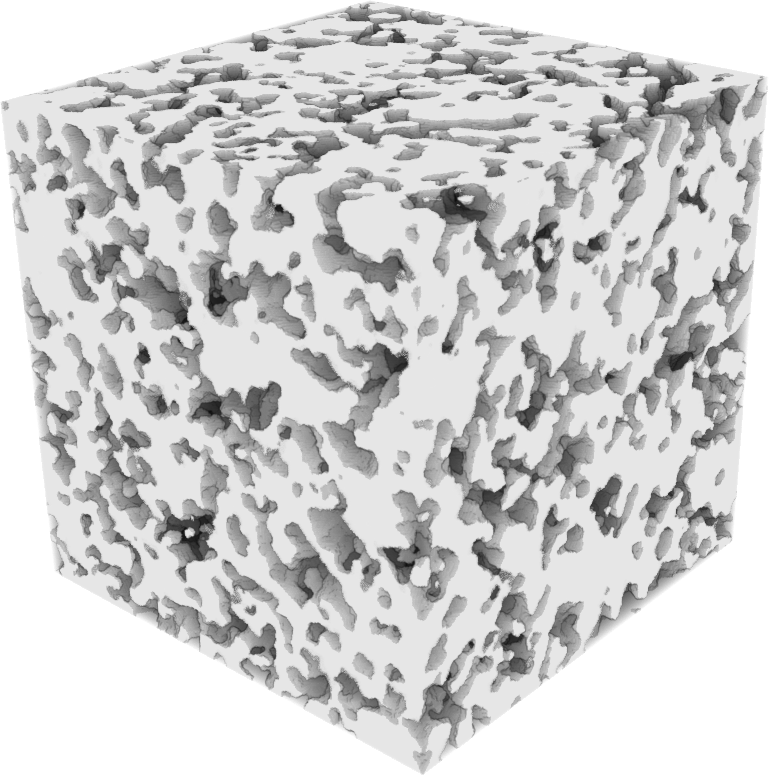}
         \caption{Ceramics SA reconstruction}
        
     \end{subfigure}
     \hfill
     \begin{subfigure}[t]{0.32\columnwidth}
         \centering
         \includegraphics[width=\textwidth]{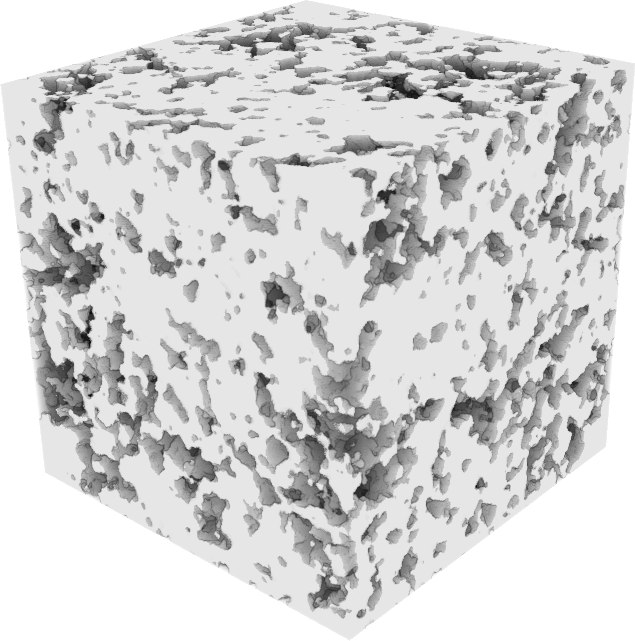}
         \caption{Ceramics PR reconstruction}
    
     \end{subfigure}
     \hfill
          \centering
     \begin{subfigure}[t]{0.32\columnwidth}
         \centering
         \includegraphics[width=\textwidth]{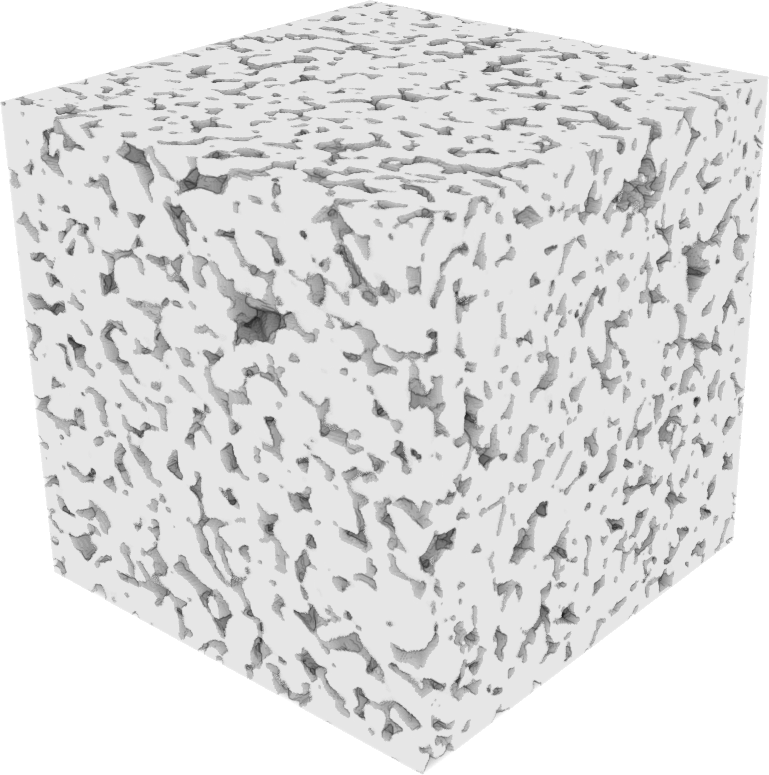}
         \caption{Sandstone original}
   
     \end{subfigure}
     \hfill
     \centering
     \begin{subfigure}[t]{0.32\columnwidth}
         \centering
         \includegraphics[width=\textwidth]{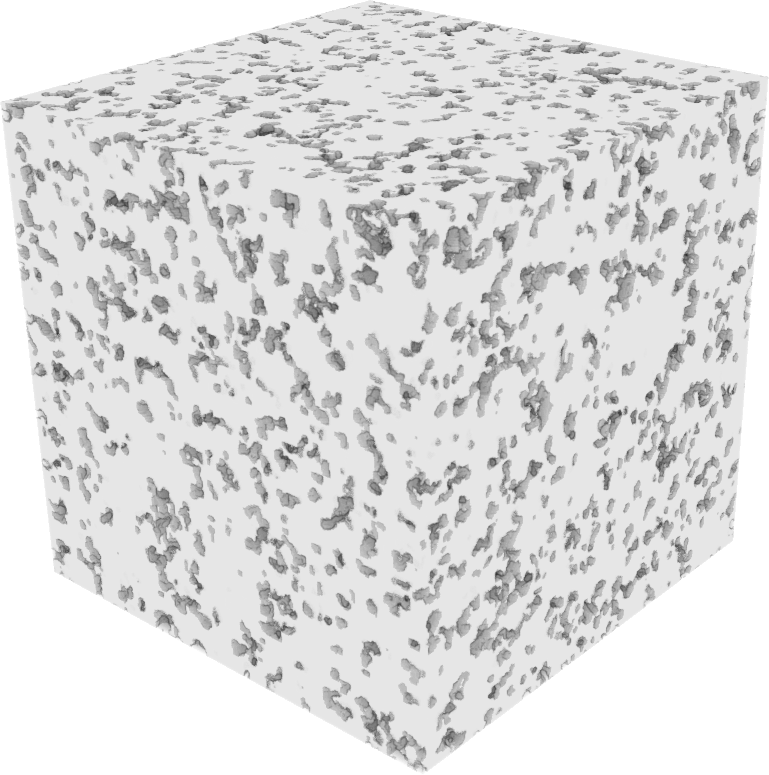}
         \caption{Sandstone SA reconstructuion}
         
     \end{subfigure}
     \hfill
     \begin{subfigure}[t]{0.32\columnwidth}
         \centering
         \includegraphics[width=\textwidth]{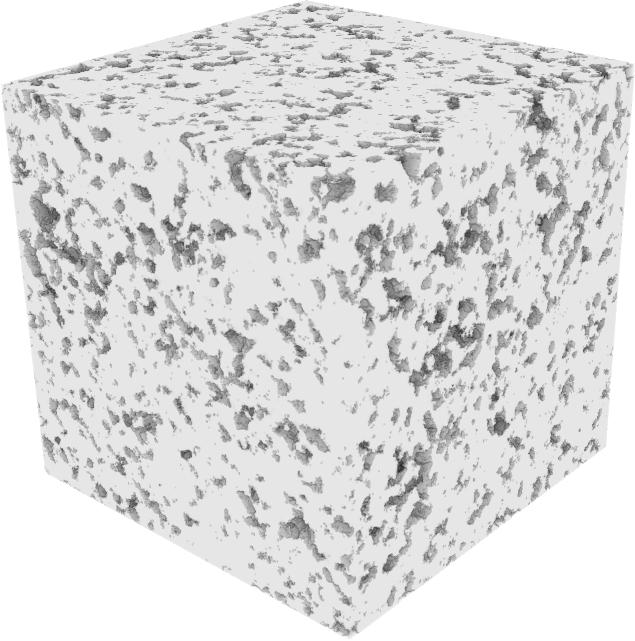}
         \caption{Sandstone PR reconstruction}
         
     \end{subfigure}
     \hfill
        \caption{Comparison of different reconstruction methods}
             \label{fig:threeMethCubes}
\end{figure}
\begin{figure}[b]
\includegraphics[width = \linewidth]{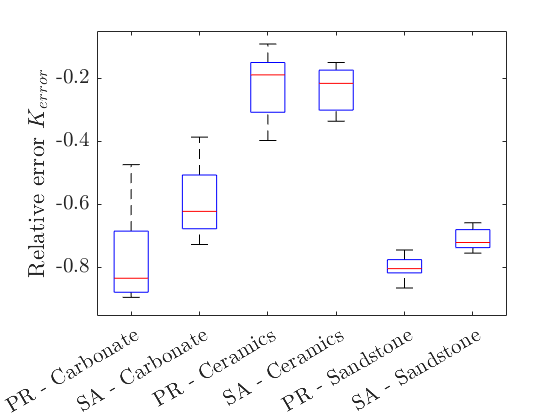}
\caption{\label{fig:bpt} Relative error $K_{error}$ comparison for carbonate, ceramics and sandstone}
\end{figure}

\begin{figure}
     \centering
     \begin{subfigure}[t]{0.9\columnwidth}
         \centering
         \includegraphics[width=\textwidth]{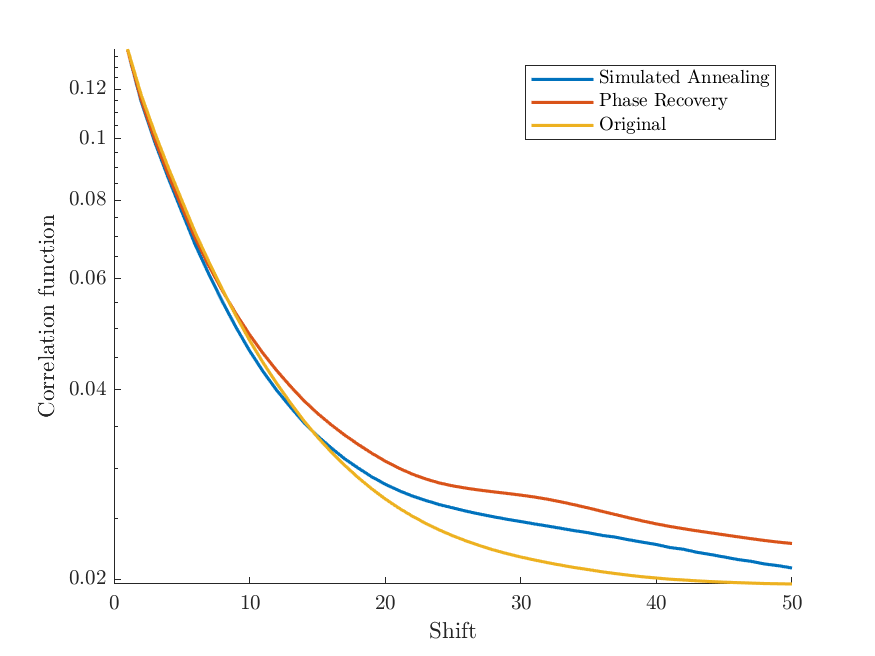}
         \caption{Correlation function}
         
     \end{subfigure}
     \hfill
     \centering
     \begin{subfigure}[t]{0.8\columnwidth}
         \centering
         \includegraphics[width=\textwidth]{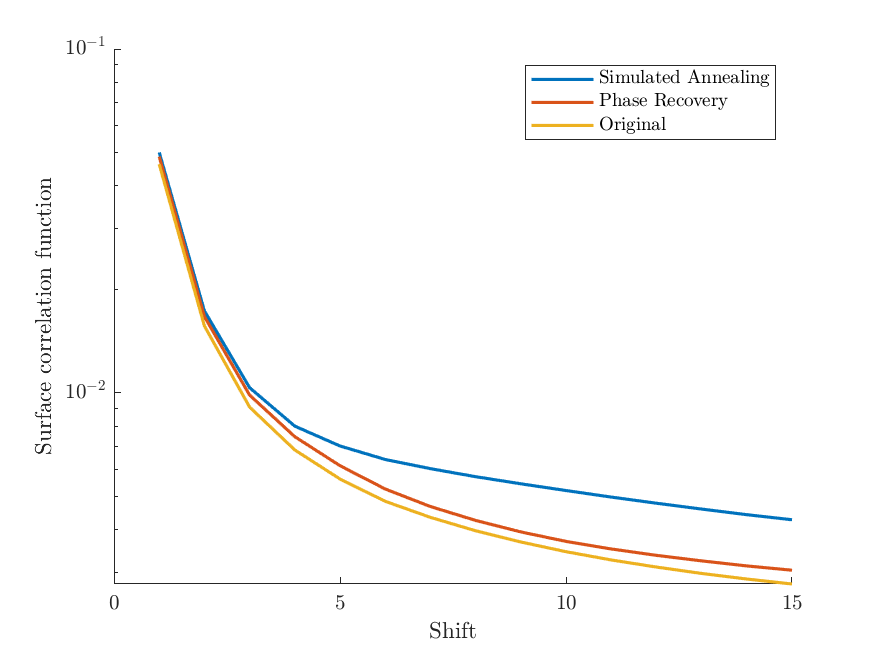}
         \caption{Surface correlation function}
         
     \end{subfigure}
     \hfill
     \\
     \centering
     \begin{subfigure}[t]{0.8\columnwidth}
         \centering
         \includegraphics[width=\textwidth]{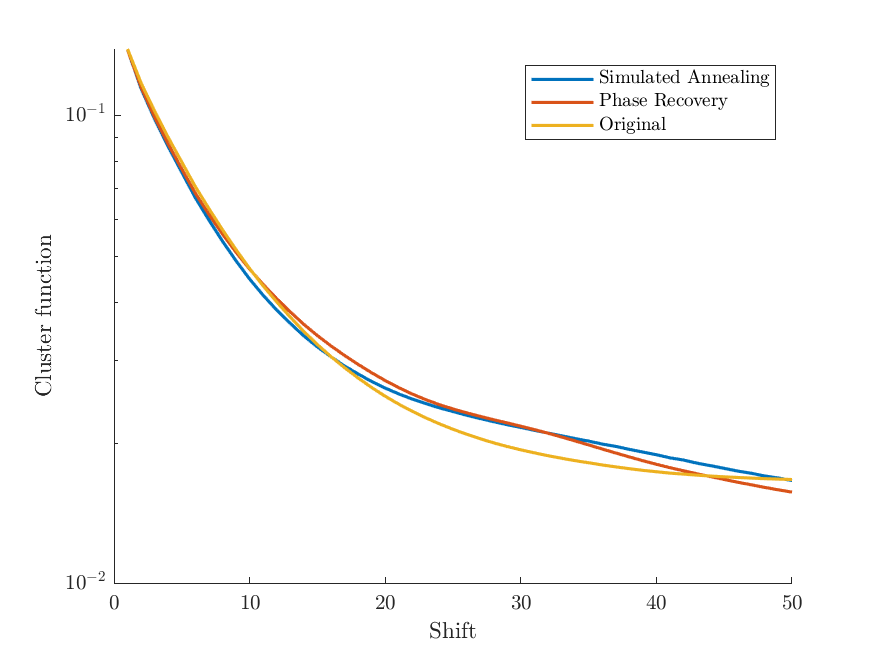}
         \caption{Cluster function}
         
     \end{subfigure}
     \hfill
     \begin{subfigure}[t]{0.8\columnwidth}
         \centering
         \includegraphics[width=\textwidth]{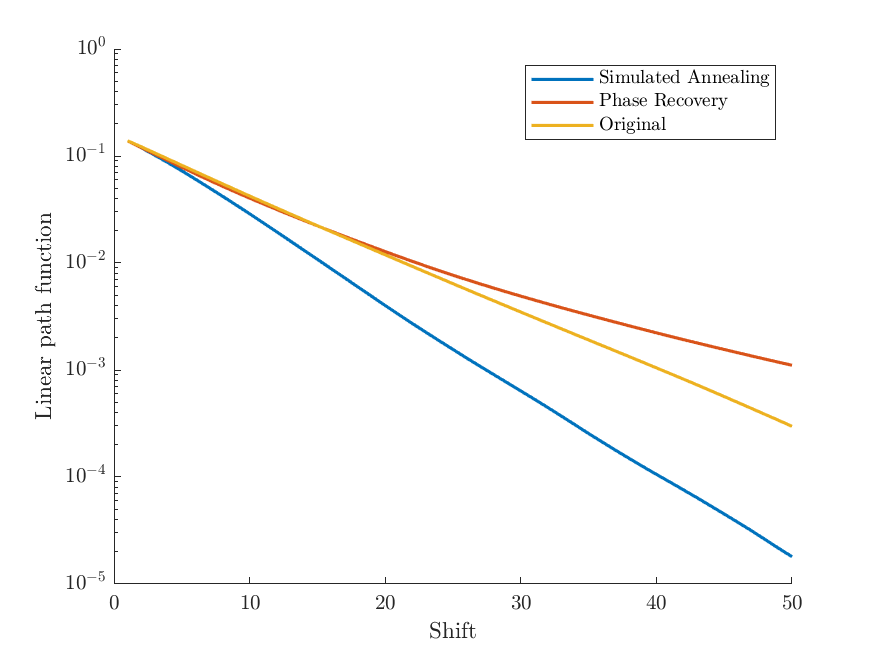}
         \caption{Linear path function}
     \end{subfigure}
     
        \caption{Comparison of quality metrics for carbonate}
        \label{fig:qualMetrCarb}
\end{figure}

\begin{figure}
     \centering
     \begin{subfigure}[t]{0.9\columnwidth}
         \centering
         \includegraphics[width=\textwidth]{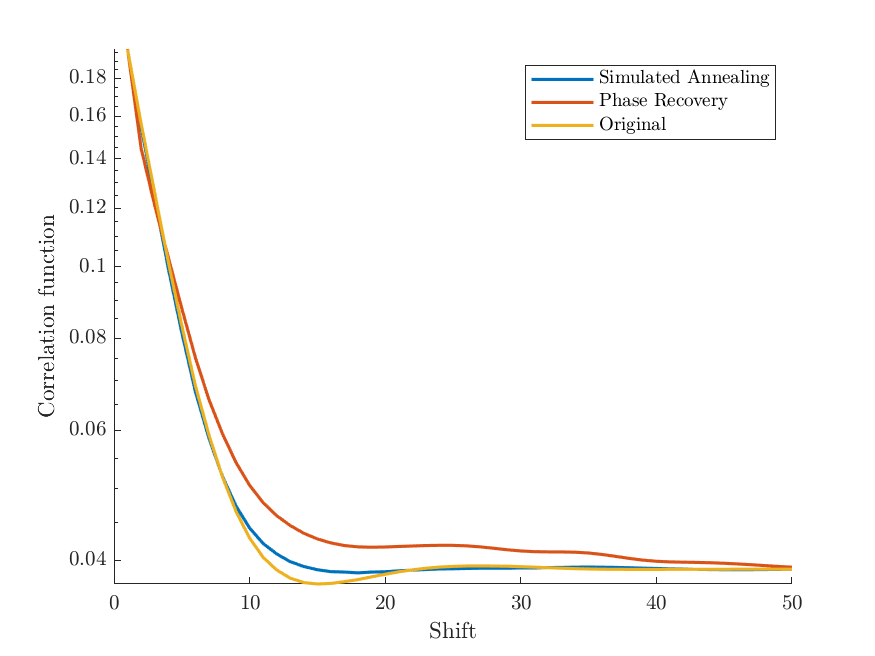}
         \caption{Correlation function}
         
     \end{subfigure}
     \hfill
     \centering
     \begin{subfigure}[t]{0.8\columnwidth}
         \centering
         \includegraphics[width=\textwidth]{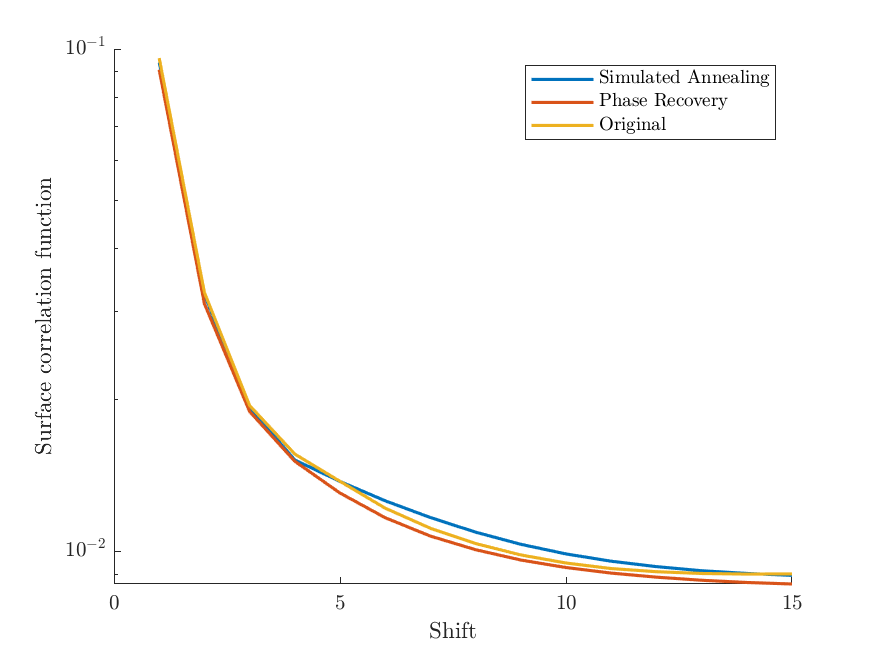}
         \caption{Surface correlation function}
         
     \end{subfigure}
     \hfill
     \\
     \centering
     \begin{subfigure}[t]{0.8\columnwidth}
         \centering
         \includegraphics[width=\textwidth]{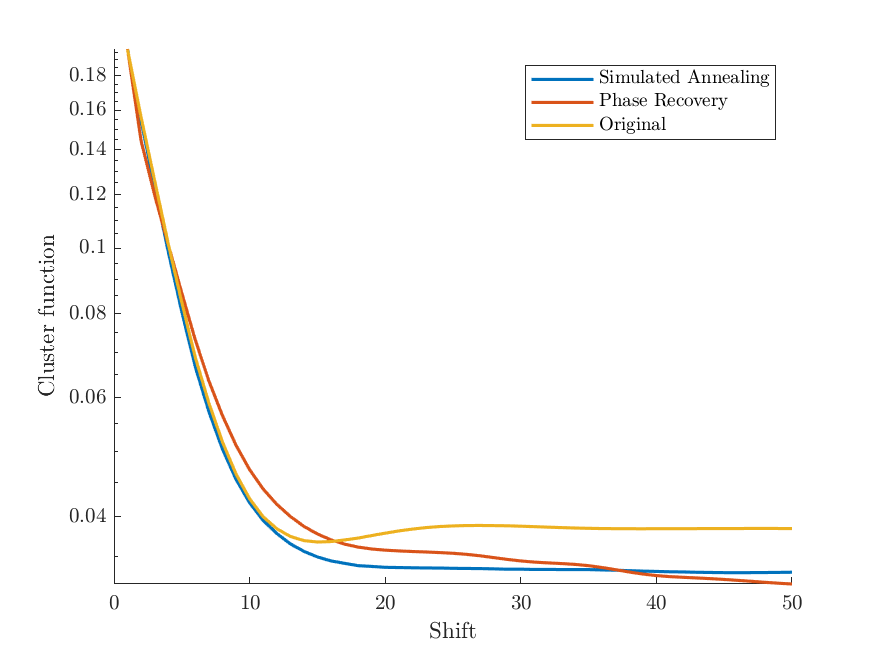}
         \caption{Cluster function}
         
     \end{subfigure}
     \hfill
     \begin{subfigure}[t]{0.8\columnwidth}
         \centering
         \includegraphics[width=\textwidth]{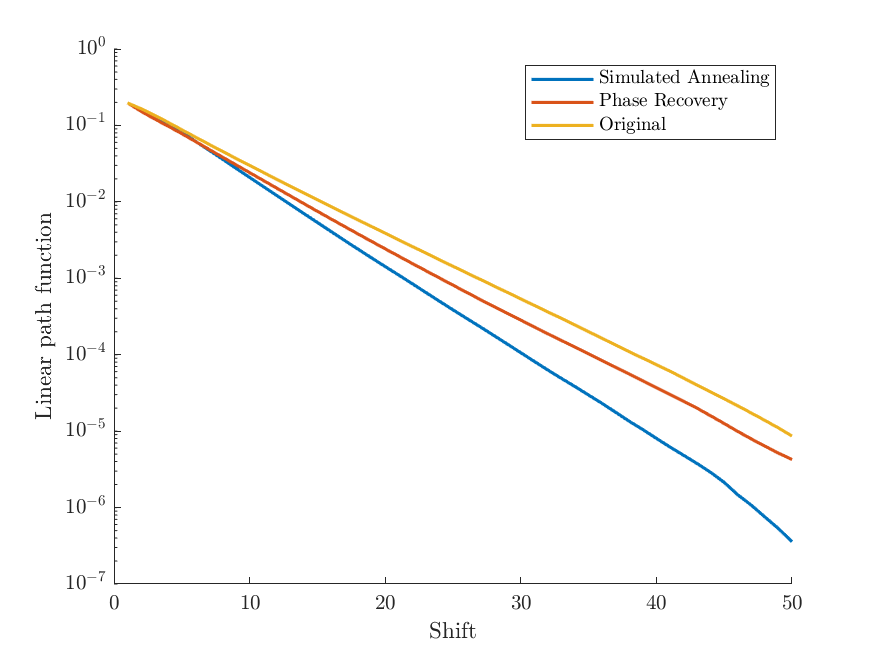}
         \caption{Linear path function}
     \end{subfigure}
     
        \caption{Comparison of quality metrics for sandstone}
        \label{fig:qualMetrSand}
\end{figure}

Visual comparison between the original binarized XCT 3D images and their reconstructed replicas revealed no particular information regarding differences between phase-retrieval and simulated annealing methods (figure \ref{fig:threeMethCubes}). To the naked eye, it seems that SA performs somewhat better for ceramic sample, while PR resembles carbonate and sandstone originals in a superior fashion. In contrast to unmodified phase-retrieval from 2D to 3D images \cite{hasanabadi2016efficient}, our approach produces no apparent artifacts or distortions. Permeabilities of the stochastic replicas provide a much more robust benchmark \ref{fig:bpt}. We immediately observe that both reconstruction methods underestimate permeability – this is to be expected as they are mainly based on the $S_2$ correlation function, which provides only a limiting information content \cite{gommes2012microstructural, gommes2012density} for complex porous media images used for the testing. For high porosity ceramic sample, PR replicas had higher permeabilities than SA, with one of the reconstructions being very similar to the original 3D image. For low porosity sandstone and carbonate samples, SA proved to provide consistently lower errors.
\\
To understand the permeability results better and to make the comparison between classical SA and developed PR method fully quantitative morphologically, further analysis in terms of $L_2$, $SS_2$, and $C_2$ correlation functions is presented in figures \ref{fig:qualMetrCarb}, \ref{fig:qualMetrSand}. For ceramic sample, PR has higher connectivity (based on $C_2$) and larger pores (based on $L_2$). This results in high permeability values as explained by large and connected pores (visible on \ref{fig:threeMethCubes}). For the carbonate sample, PR outperforms SA based on all additional CFs. Correlation functions for sandstone replicas reveal no obvious winner, as phase-retrieval shows better $L_2$ statistics while diverging more compared to annealing in terms of $C_2$. This overall situation is rather interesting, as PR replicas being better in terms of CFs did not guarantee their permeability error $K_{error}$ lower that for SA reconstructions. Note that the $S_2$ correlation function's match was always perfect by design (both PR and SA minimized the difference).
\\

\begin{figure}
     \centering
     \begin{subfigure}[t]{0.9\columnwidth}
         \centering
         \includegraphics[width=\textwidth]{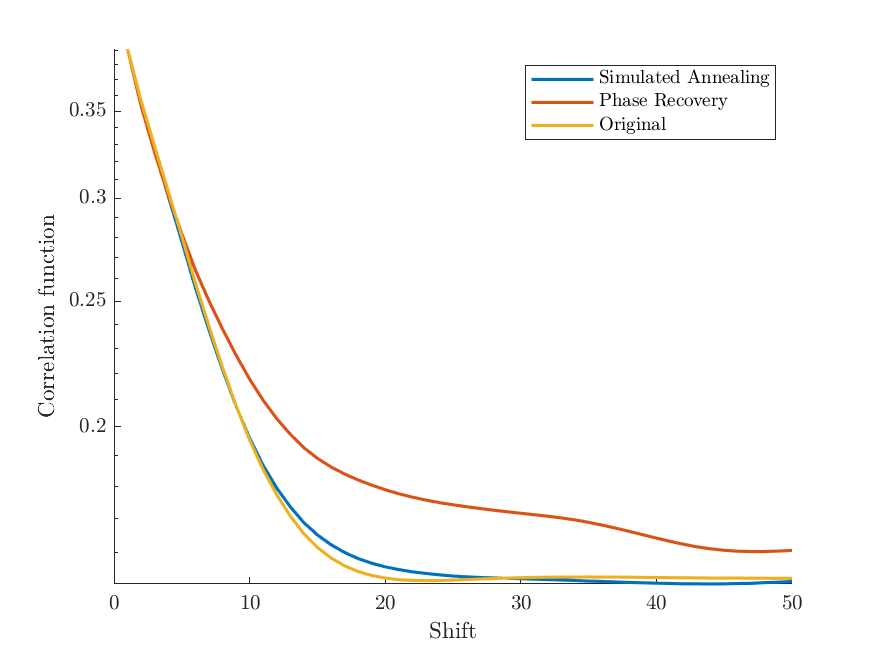}
         \caption{Correlation function}
         
     \end{subfigure}
     \hfill
     \centering
     \begin{subfigure}[t]{0.8\columnwidth}
         \centering
         \includegraphics[width=\textwidth]{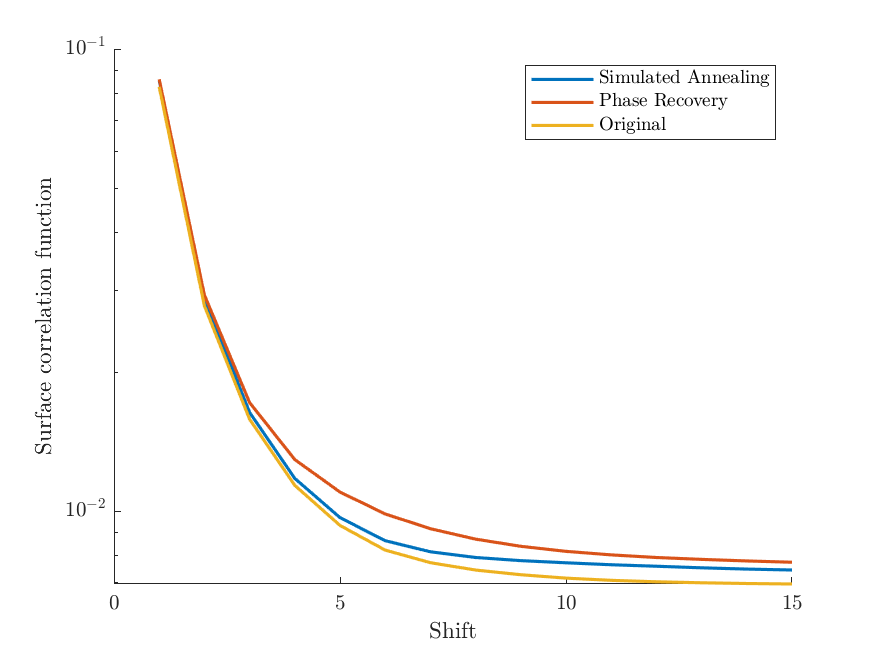}
         \caption{Surface correlation function}
         
     \end{subfigure}
     \hfill
     \\
     \centering
     \begin{subfigure}[t]{0.8\columnwidth}
         \centering
         \includegraphics[width=\textwidth]{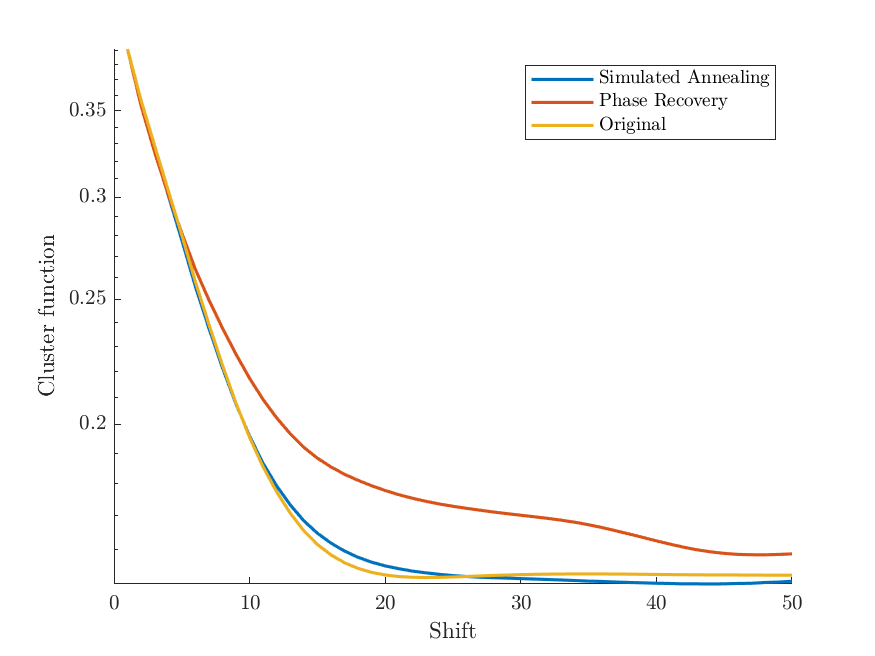}
         \caption{Cluster function}
         
     \end{subfigure}
     \hfill
     \begin{subfigure}[t]{0.8\columnwidth}
         \centering
         \includegraphics[width=\textwidth]{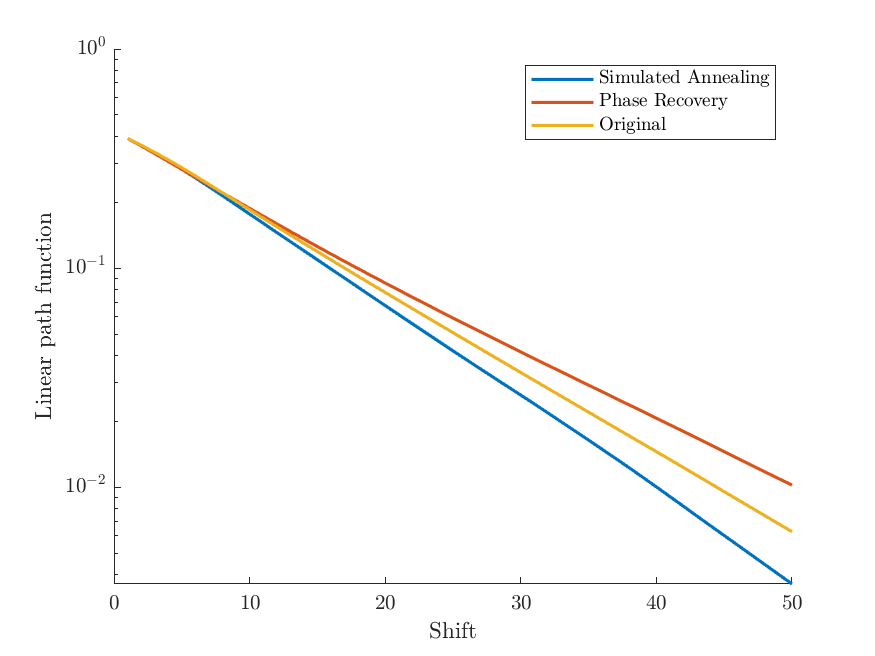}
         \caption{Linear path function}
     \end{subfigure}
     
        \caption{Comparison of quality metrics for ceramics}
        \label{fig:qualMetrCeram}
\end{figure}

Putting all the results from \ref{fig:threeMethCubes}-\ref{fig:qualMetrSand} together, it is impossible to choose either PR or SA as a more favorable approach, and this was not the target of the comparison. All in all, we conclude that both approaches are comparable in terms of their accuracy and perform differently for various sample structures. We iteratively improve the structure by voxel permutations with annealing, while phase-retrieval is more of a sequential algorithm. As SA swaps voxels preferentially to the interfaces between binary phases for improved convergence \cite{tang2009pixel, vcapek2009stochastic, pant2014stochastic}, we introduced surface-surface optimization \cite{ma2018precise} within PR to counterbalance such permutations. Without this augmentation, i.e., if based on the $S_2$ correlation function alone, PR performed slightly worse than SA. Thus, the current version of the comparison is deemed fair. Based on the large variety of images used in this work, we are safe to conclude that annealing with $S_2$ computed in orthogonal and diagonal directions produces consistently better results as compared to full-map $S_2$ self-convolution and, thus, annealing can be considered to be a polishing step \cite{politis2008hybrid}.
\\
While classic SA and modified PR method as described here do provide quite similar reconstructions in terms of balanced morphological and permeability assessment, phase-retrieval is computationally much less taxing. Based on our Matlab implementation, phase-retrieval with 40 optimization steps with 20 iterations each took approximately 40 mins for $300^3$ voxels replicas. In contrast, simulated annealing took around 2 h for the same problem. While comparison of SA-based reconstruction times is always hampered by stopping criteria, these two 2 h was the average time needed to overcome PR in accuracy based on $S_2$ correlation statistics used as optimization targets in both algorithms. We believe that our PR implementation has much potential for speeding up and cannot be compared directly against legacy C++ implementation (the SA method).
\\
Another interesting aspect of the DDTF approach is its ability to utilize transition functions (\ref{fig:dimef}) to create stochastic yet highly deterministic 3D structures from 2D images. For example, this ability could be used to re-create 3D grain morphology from 2D cuts. This way, phase-retrieval can be coupled with process-based reconstructions \cite{oren2002process, thovert2011grain} to create three-dimensional shapes during the packing process.\\
Most importantly, in this work, we have explicitly shown that phase-retrieval can utilize more sophisticated structural statistics beyond the $S_2$ two-point correlation function - we tried $SS_2$ and $C_2$ so far, or other morphological metrics such as fractal dimensions \cite{khlyupin2015fractal, ju20143d}. This means that fast 3D reconstructions based on modified PR with the incorporation of other additional correlation functions can be used as input data for the SA algorithm \cite{politis2008hybrid} or building block for MPS reconstruction \cite{tahmasebi2013cross, tahmasebi2012reconstruction, gravey2020quicksampling, tahmasebi2013cross}, as well as the abovementioned hybridization with process-based methods. Such potential hybridization of different methods and our modification that makes phase-retrieval similar to other main-stream reconstruction techniques paves the way to the overall unification of all approaches, as they solve the very similar optimization problem.\\
\section{Conclusions}
In this work, we developed a dynamic phase-retrieval stochastic reconstruction algorithm for creating 3D replicas from 2D original mages - DDTF. The DDTF is free of artifacts characteristic of previously proposed phase-retrieval techniques. While based on two-point $S_2$ correlation functions, any correlation function (surface-surface correlation was utilized here) or other morphological metrics can be accounted for during the reconstruction, thus, paving the way to the hybridization of different reconstruction techniques. To test DDTF, we performed reconstructions for three binary porous media samples of different genesis. Based on computed permeability and connectivity ($C_2$ and $L_2$ correlation functions), we have shown that the proposed technique in terms of accuracy is comparable to the classic simulated annealing-based reconstruction method but is computationally very effective. This opens the possibility of utilizing DDTF to produce fast or crude replicas further improved by other reconstruction techniques such as SA or process-based methods. Improving the quality of reconstructions based on phase-retrieval by adding additional metrics into the reconstruction procedure is a research activity with a high potential to enhance the methodology significantly.

\begin{acknowledgments}
This work was supported by Russian Science Foundation grant №19-72-10082 (M.V.K. and K.M.G.). The generous cooperation of the authors is within the Flow and Transport in Media with Pores research group (FaT iMP, www.porenetwork.com) and uses some of its software.
\end{acknowledgments}

\appendix

\section{Maximal likelihood}
The way to estimate probability that certain discredized curve $y_{certatain}(x_i)$ belongs to set of curves $y_j(x_i), ..., y_N(x_i)$ is the maximal likelihood method. 
The cost objective $\delta$ needed to be minimized in this case is the following.
\begin{equation}
    \delta = \sum_i{((\frac{y(x_i) - y_{mean}(x_i)}{\sigma_i})^2 + \ln(2\pi \sigma_i))}
    \label{eq:lkhd}
\end{equation}
where mean $y_{mean}$ and standard deviation $\sigma_i$ are defined in \ref{eq:mean} and \ref{eq:sigma} respectively.
\begin{equation}
   y_{mean}(x_i) = \frac{1}{N}\sum_j{y_{j}(x_i)}
   \label{eq:sigma}
\end{equation}

\begin{equation}
   \sigma_{i}=\sqrt{ \frac{1}{N}\sum_j{(y_{j}(x_i)} - y_{mean}(x_i))^2}
   \label{eq:mean}
\end{equation}

\nocite{*}

\bibliography{apssamp}

\end{document}